\documentclass[useAMS,usenatbib]{mn2e}
\usepackage{graphicx,natbib}
\usepackage{psfrag}
\usepackage{amssymb}
\usepackage{float}
\usepackage[dvips]{color}
\usepackage[export]{adjustbox}  
\usepackage{auto-pst-pdf}
\usepackage{pstricks}
\usepackage{pst-all}
\usepackage{multicol}

\usepackage{supertabular}

\usepackage{changepage}
\usepackage{pdflscape}
\usepackage{lmodern}
\usepackage{lscape}
\usepackage{longtable}
\usepackage{dcolumn}

\title[WISE embedded cluster survey]{Characterizing star cluster formation with WISE: 652 newly found star clusters and candidates}

\author[D. Camargo, E. Bica, and C. Bonatto]{D. Camargo$^{1}$,  E. Bica$^2$, and C. Bonatto$^2$\\
$^1$ Col{\'e}gio Militar de Porto Alegre, Minist{\'e}rio da Defesa - Ex{\'e}rcito Brasileiro, 
Av. Jos{\'e} Bonif{\'a}cio 363\\
Porto Alegre 90040-130, RS, Brazil\\
$^2$ Departamento de Astronomia, Universidade Federal do Rio Grande do Sul, 
Av. Bento Gon\c{c}alves 9500\\
Porto Alegre 91501-970, RS, Brazil}

\begin{document}

\pagerange{\pageref{firstpage}--\pageref{lastpage}}

\maketitle

\label{firstpage}

\begin{abstract}
We report the discovery of 652 star clusters, stellar groups and candidates in the Milky Way with WISE. Most of the objects are projected close to Galactic Plane and are embedded clusters. The present sample complements a similar study (Paper I)  which provided 437 star clusters and alike. We find evidence that star formation processes span a wide range of sizes, from populous  dense clusters to small compact embedded ones, sparse stellar groups or in relative isolation.
The present list indicates multiple stellar generations during the embedded phase, with giant molecular clouds collapsing into several clumps composing an embedded cluster aggregate. We investigate the field star decontaminated Colour Magnitude Diagrams and Radial Density Profiles of 9 cluster candidates in the list, and derive their parameters, confirming them as embedded clusters.
\end{abstract}


\section{Introduction}
\label{Intro}

Star cluster formation is a complex process, which is not yet completely understood. These systems are formed within giant molecular clouds (GMCs) that populate the Galactic disc, mainly the spiral arms in an early evolutionary stage. They are known as embedded clusters (ECs) and/or embedded stellar groups (EGrs). The molecular clouds, in which ECs are formed often present filamentary structures \citep{Gutermuth09, Myers09, Camargo12} that also may be the site of small EGr formation \citep{Bastian11, Camargo12, Kruijssen12}. 

Thus, ECs are fundamental tools to improve our understanding of star formation and evolution as well as the Galactic structures that trace the spiral pattern \citep{Camargo13, Camargo15c}. These objects also play important roles in both local and large scale star formation clarifying the physical processes involved in the individual and Galactic scale of GMCs. 

Given the importance of embedded clusters as tools to study the Galactic structure, several surveys have been recently carried out \citep{Bica03a, Bica03b, Lada03, Bica05, Borissova11, Majaess13, Borissova14, Camargo15a}, together with a detailed analysis of their properties \citep[e.g.][]{Camargo10, Camargo11, Camargo12, Camargo13}. 

These surveys suggest that most stars form in clumped environments, with scales ranging from small stellar groups to massive clusters or large associations \citep{Cotera99, Motte01, Mauerhan10, Kirk12, Oskinova13, Habibi14}. Infrared surveys, such as the Two Micron All Sky Survey (2MASS), Spitzer, and WISE are uncovering deep embedded clusters and embedded stellar groups in the Galaxy.

\begin{figure*}
\begin{minipage}[b]{0.328\linewidth}
\includegraphics[width=\textwidth]{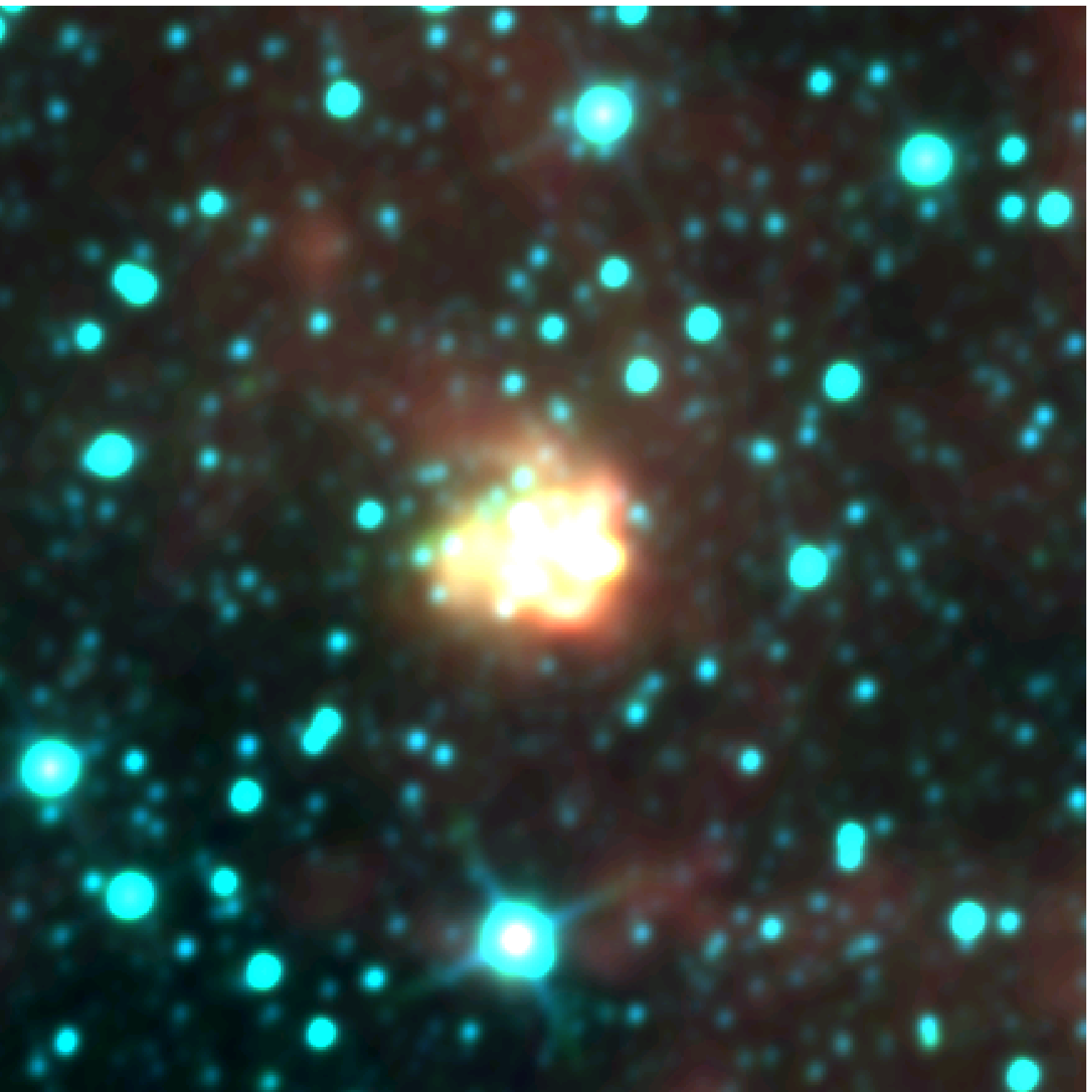}
\put(-134.0,155.0){\makebox(0.0,0.0)[5]{\fontsize{14}{14}\selectfont \color{red} C 791}}
\end{minipage}\hfill
\vspace{0.02cm}
\begin{minipage}[b]{0.328\linewidth}
\includegraphics[width=\textwidth]{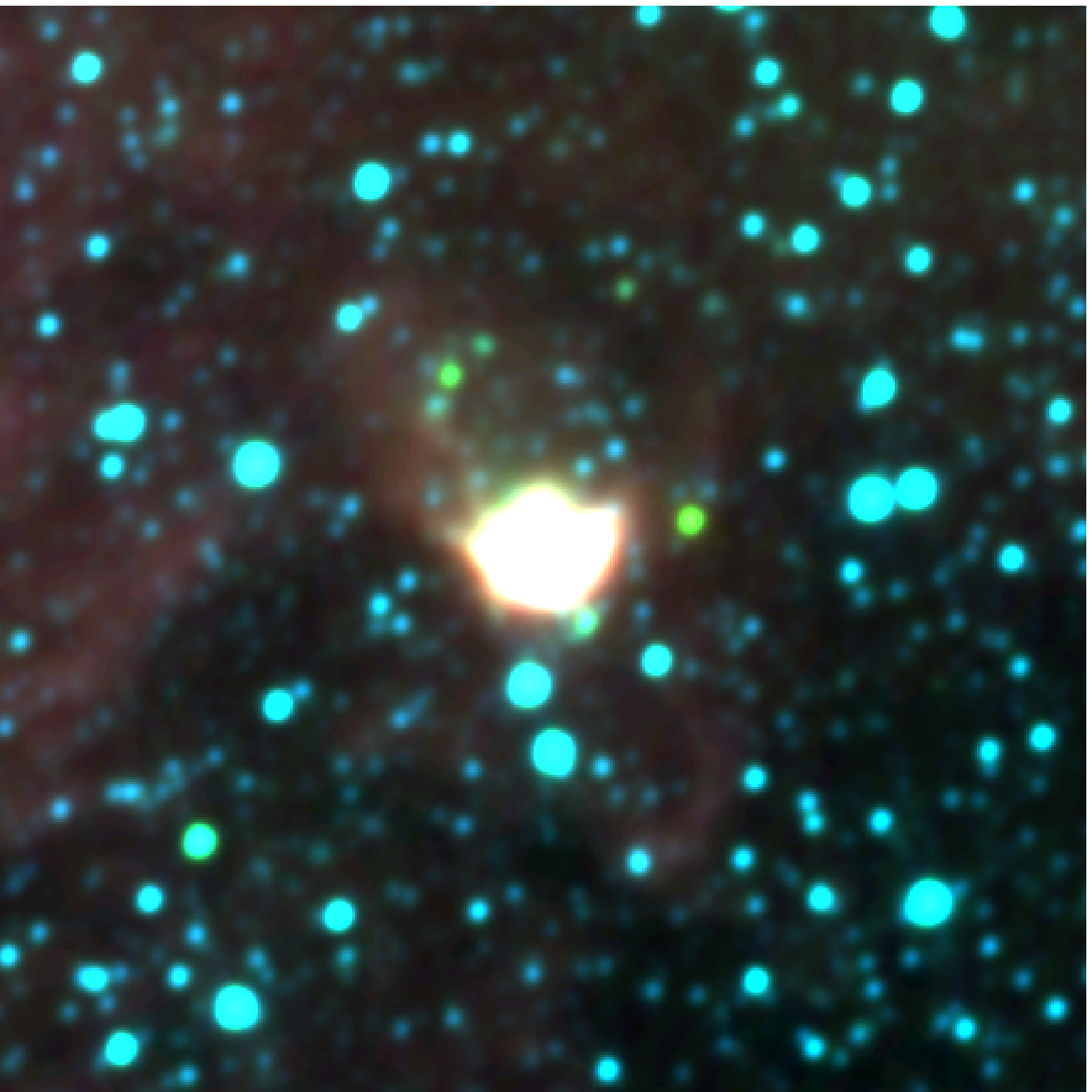}
\put(-140.0,155.0){\makebox(0.0,0.0)[5]{\fontsize{14}{14}\selectfont \color{red} C 788}}
\end{minipage}\hfill
\vspace{0.02cm}
\begin{minipage}[b]{0.328\linewidth}
\includegraphics[width=\textwidth]{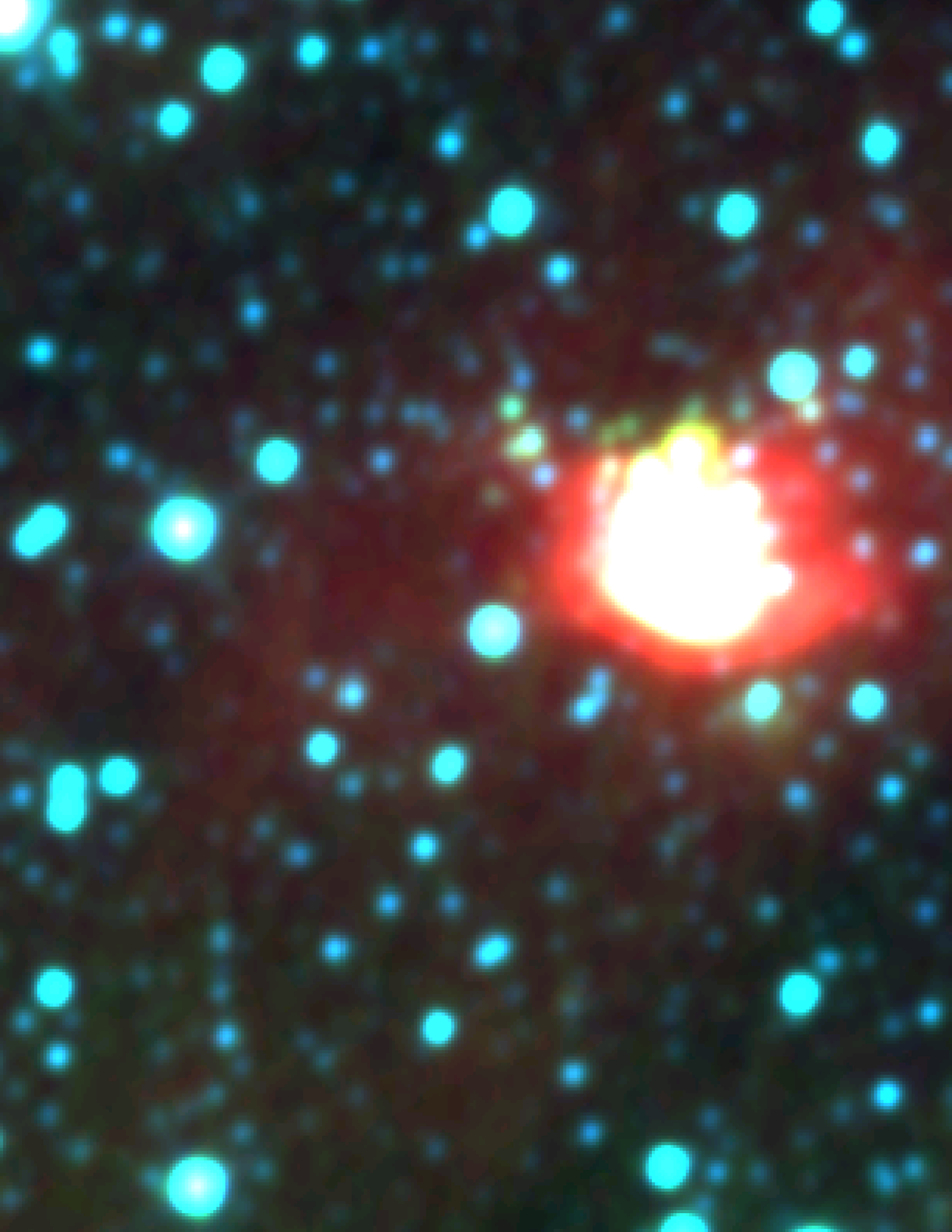}
\put(-140.0,155.0){\makebox(0.0,0.0)[5]{\fontsize{14}{14}\selectfont \color{red}C 860}}
\end{minipage}\hfill
\vspace{0.02cm}
\begin{minipage}[b]{0.328\linewidth}
\includegraphics[width=\textwidth]{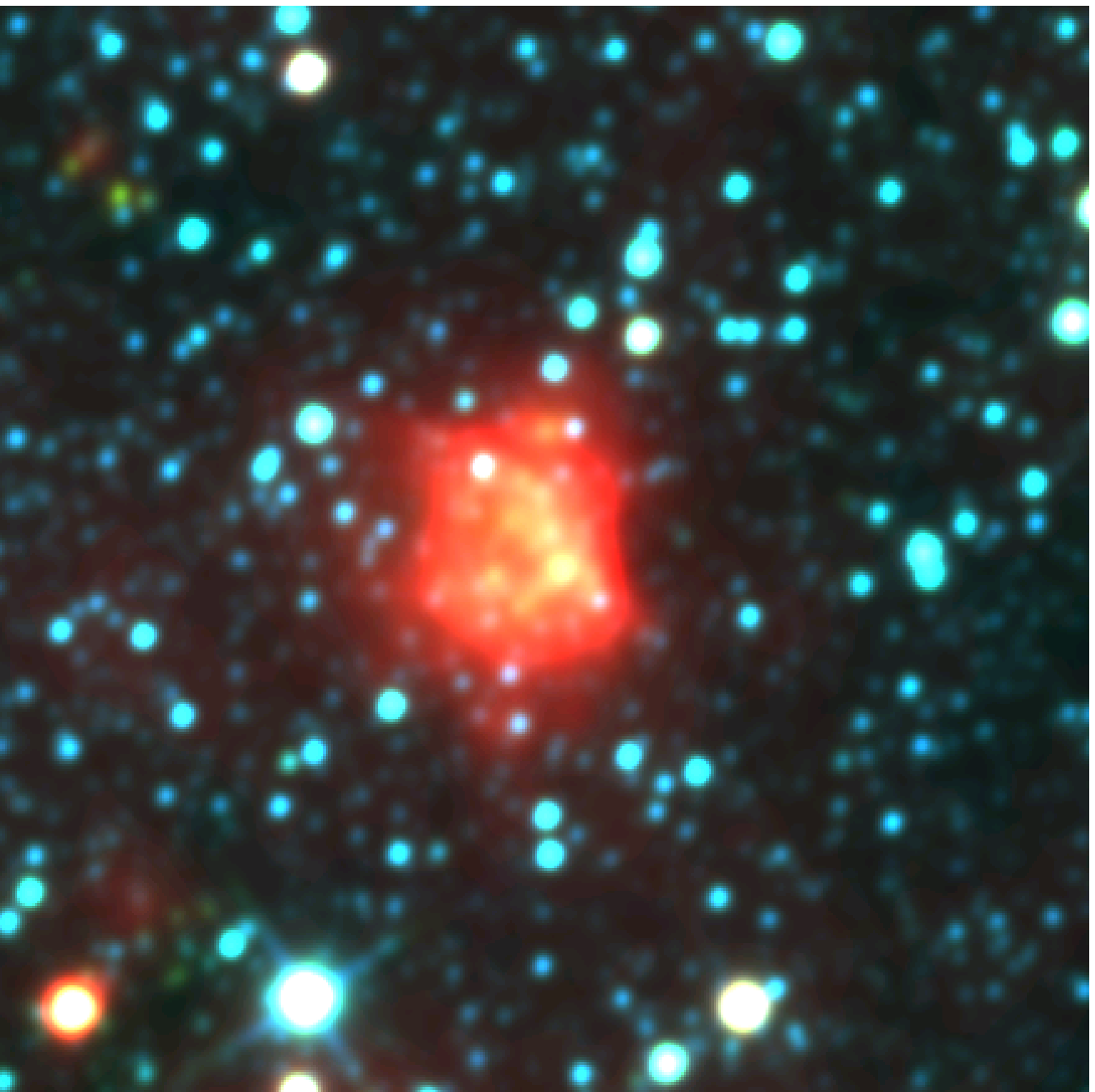}
\put(-140.0,155.0){\makebox(0.0,0.0)[5]{\fontsize{14}{14}\selectfont \color{red}C 514}}
\end{minipage}\hfill
\begin{minipage}[b]{0.328\linewidth}
\includegraphics[width=\textwidth]{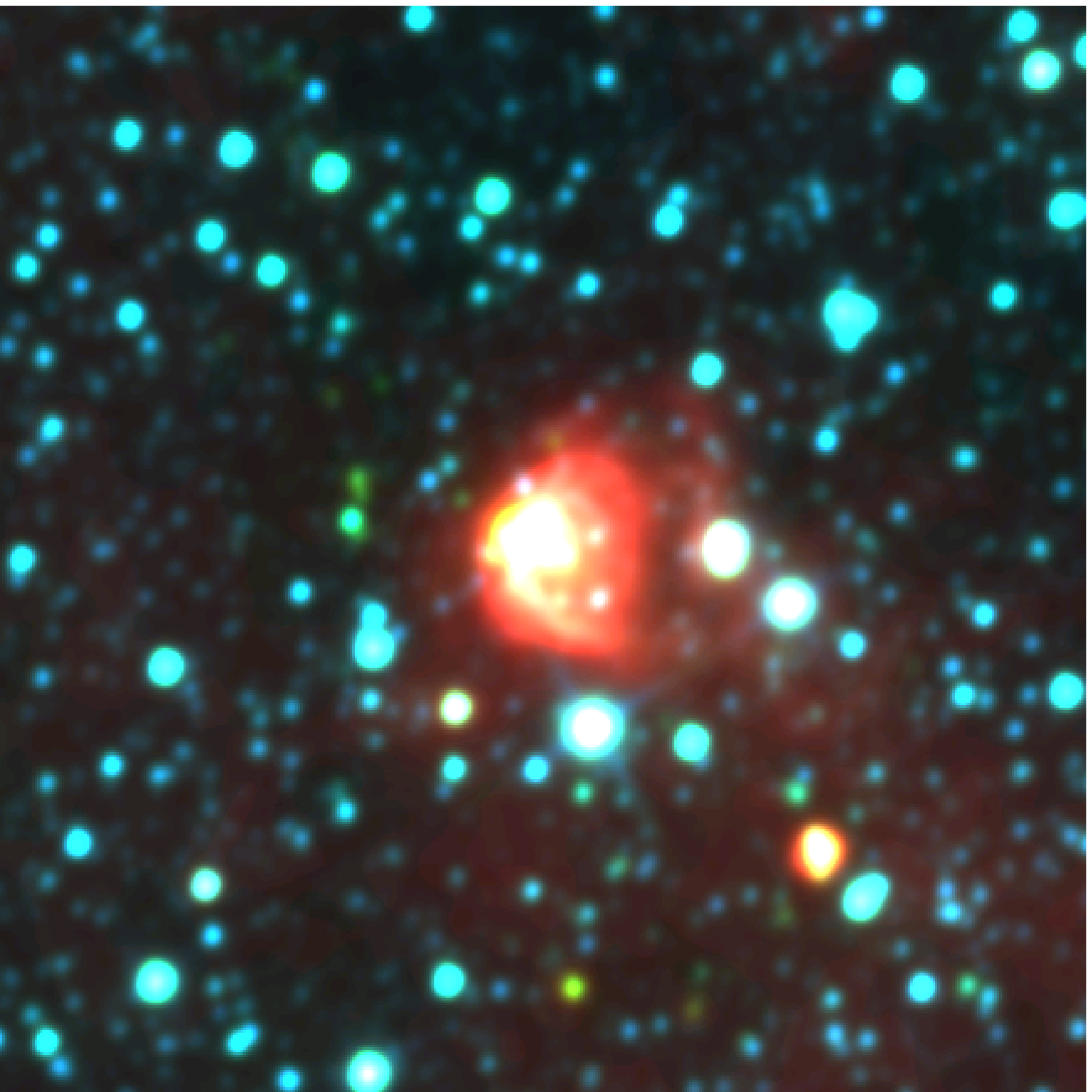}
\put(-140.0,155.0){\makebox(0.0,0.0)[5]{\fontsize{14}{14}\selectfont \color{red}C 530}}
\end{minipage}\hfill
\begin{minipage}[b]{0.328\linewidth}
\includegraphics[width=\textwidth]{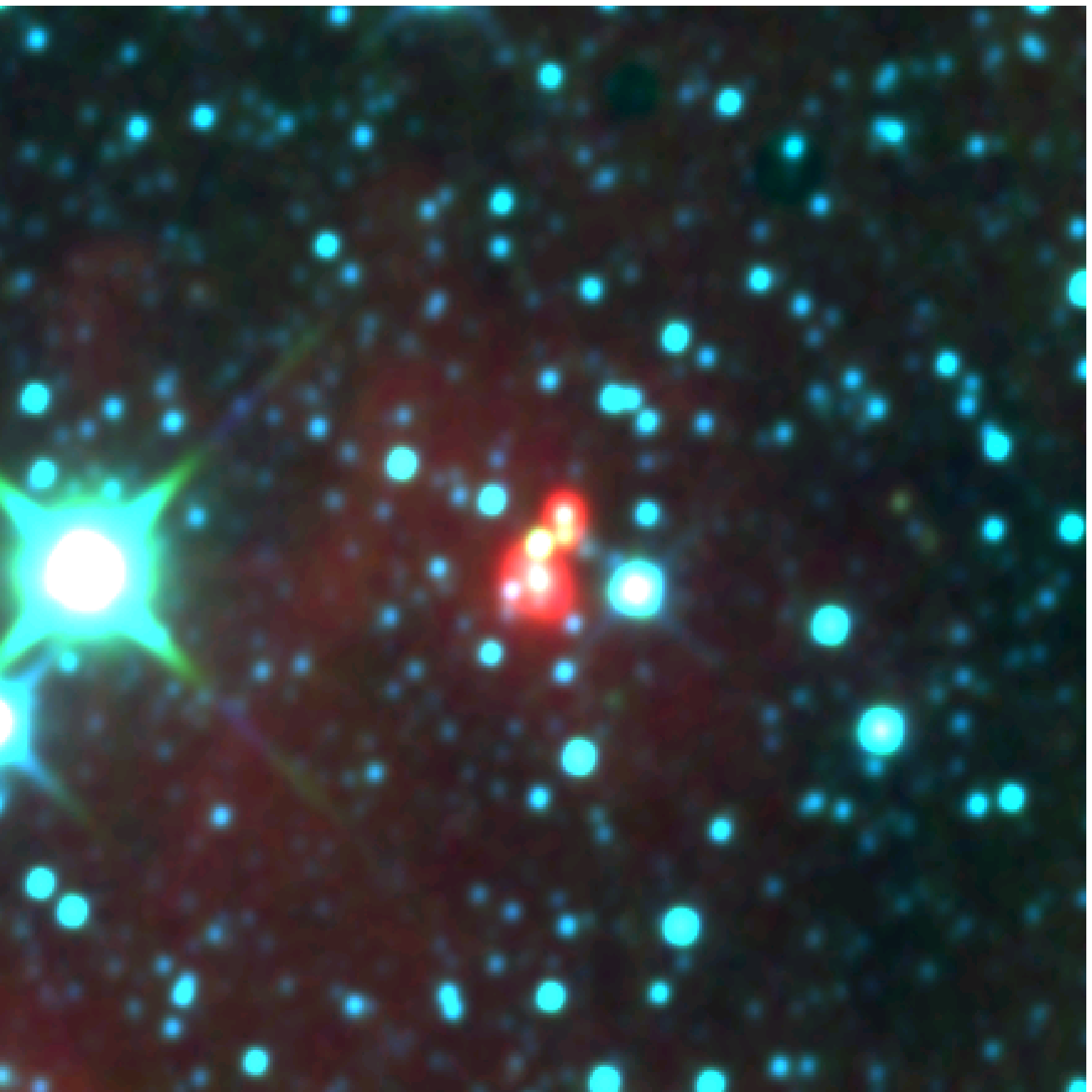}
\put(-140.0,155.0){\makebox(0.0,0.0)[5]{\fontsize{14}{14}\selectfont \color{red}C 941}}
\end{minipage}\hfill
\caption[]{WISE ($10'\times10'$) RGB images centred on the embedded clusters C 791, C 788, C 860, C 514, C 530, and C 941.}
\label{f1}
\end{figure*}

Despite advances in our understanding of star cluster formation and their early evolution, there is no consensus on the definition of an embedded cluster. There is evidence that in the early evolutionary stages some embedded clusters may contain less than $\sim10$ stars and radii of about 0.2 pc \citep{Testi97, Massi06, Rodrigues08, Alexander12}. In this sense, \citet{Hodapp94} define clusters as groups of five or more stars. \citet{Adams01} classify as stellar groups systems with 10 to 100 stars, and the most populous ones as clusters. \citet{Lada03} suggest that at least 35 stars and stellar mass density of about $1M_{\odot}/pc^{-3}$ \citep{Bok34, Spitzer58} are required for an EC to survive the tidal disruption by the Galaxy, GMC collisions, and the gas expulsion as a bound system. However, as they suggest that most ECs do not survive the \textit{infant mortality}, this criterion is only a lower limit for ECs to survive the gas expulsion to become an open cluster. This is not a criterion for EC classifications, on the contrary, it shows that most ECs are below this limit since they argue that more than $95\%$ of them are dissolved before 100 Myr.

\begin{table*}
\centering
\caption{List of newly found star clusters and candidates.}
\label{tab1}
\renewcommand{\tabcolsep}{3.6mm}
\renewcommand{\arraystretch}{1.2}
\begin{tabular}{lrrrrrrrr}
\hline
\hline
Target&$\ell$&$b$&$\alpha(2000)$&$\delta(2000)$&Size&Type&Comments\\
&$(^{\circ})$&$(^{\circ})$&(h\,m\,s)&$(^{\circ}\,^{\prime}\,^{\prime\prime})$&$(\,^{\prime}\,)$&&\\ \hline
C 514 &56.15&   0.08& 19:35:25&  20:34:49&$  3.5\times  3.5$&  EC&     impressive dusty core\\
C 530 &58.77&   0.65& 19:38:48&  23:08:31&$    6\times    6$&  EC&        impressive, dusty core, include bubble\\
C 532 &58.95&   1.39& 19:36:22&  23:39:46&$    5\times    5$&  EGr&         protocluster?\\
C 578 &64.38&  -0.34& 19:55:07&  27:29:35&$    6\times    6$&  EGr&      include IR Dark and IR Bright Nebula\\ 
C 582 &64.96&  -0.04& 19:55:17&  28:08:47&$    3\times    3$&  OCC& \\         
C 649 &119.72&   1.68&  0:21:42&  64:21:58&$    6\times    4$&  EC&        prominent, in pair with King 1 \\
C 700 &127.02&   0.76&  1:28:00&  63:20:42&$    4\times    3$&  ECC& \\
C 716  &131.45&  -1.16&  2:01:31&  60:32:42&$    6\times    6$&  EC&         prominent \\ 
C 741  &134.02&   0.40&  2:25:39&  61:13:33&$    6\times    6$&  EC &             small dusty stellar core \\
C 758  &135.35&   0.94&  2:37:33&  61:13:21&$    5\times    5$&  ECC&             dusty, poor, protocluster? \\
C 769  &136.38&   0.68&  2:44:24&  60:34:16&$    6\times    6$&  EC &         small stellar core\\
C 788  &139.96&   2.59&  3:17:28&  60:32:24&$    6\times    6$&  EC &         dusty core\\
C 789  &140.05&   1.91&  3:15:11&  59:54:44&$    5\times    5$&  EC &          prominent\\
C 791  &140.21&   2.54&  3:18:57&  60:22:03&$    6\times    6$&  EC &        small dusty and stellar core\\
C 793  &140.22&   2.16&  3:17:23&  60:02:07&$    7\times    7$&  EC &        prominent \\
C 836  &150.05&   3.07&  4:18:50&  54:37:59&$    6\times    6$&  OC&    \\
C 838  &150.11&   0.73&  4:08:14&  52:53:42&$    6\times    6$&  EC&  impressive stellar dominated core\\
C 853  &151.63&  -1.10&  4:07:32&  50:30:48&$    6\times    6$&  EC &       impressive, dusty stellar core \\
C 860  &151.97&  -1.97&  4:05:27&  49:38:23&$    5\times    5$&  EC &    impressive dusty core \\
C 911  &172.07&   2.32&  5:34:56&  36:53:22&$    5\times    5$&  EC &        stellar core\\
C 914  &172.73&   2.49&  5:37:24&  36:25:28&$    5\times    5$&  EC &                  \\ 
C 915  &172.76&   2.10&  5:35:53&  36:10:23&$    5\times    5$&  EC &                \\
C 916  &172.77&   2.37&  5:37:02&  36:19:33&$    5\times    5$&  EC &         stellar core\\
C 919  &173.01&   2.38&  5:37:40&  36:07:44&$    4\times    4$&  EC &        chain of stars \\ 
C 921  &173.16&   2.56&  5:38:51&  36:06:09&$    4\times    4$&  EC &       \\
C 925  &173.33&   2.37&  5:38:30&  35:50:46&$    5\times    5$&  EC &\\
C 935  &189.03&   2.91&  6:16:43&  22:32:15&$    7\times    4$&  EC &     substructured, in L dust emission arc\\                      
C 936  &189.18&   2.87&  6:16:54&  22:22:58&$    9\times    9$&  EC &     substructured, in L dust emission arc \\                                        
C 937  &189.25&   3.16&  6:18:07&  22:27:39&$    8\times    5$&  EC &     substructured, in L dust emission arc \\      
C 938  &189.29&   3.02&  6:17:42&  22:21:30&$    6\times    3$&  EC &      in large dust emission arc \\ 
C 941  &191.51&  -0.60&  6:08:39&  18:40:27&$    5\times    5$&  EC &      impressive\\ 
C 943  &191.97&   0.97&  6:15:25&  19:01:40&$  2.5\times  2.5$&  EC &       not FSR 909\\ 
C 978  &197.14&  -3.10&  6:10:50&  12:32:45&$    6\times    6$&  EC  &      prominent \\
C 985  &236.59&  -2.30&  7:27:58& -22:02:03&$    6\times    6$&  EC  &       impressive \\
C 1008  &238.43&  -4.12&  7:24:44& -24:30:45&$    7\times    7$&  EC &       in group\\                                        
C 1009  &238.47&  -4.17&  7:24:40& -24:34:24&$    4\times  3.5$&  EC &       in group\\
C 1043  &244.92&  -7.20&  7:26:04& -31:39:24&$    6\times    6$&  EC &      impressive\\
C 1086  &249.24&  -5.06&  7:44:49& -34:24:40&$    6\times    6$&  EC  &   impressive, stellar compact core, shock \\ 
\hline
\end{tabular}
\begin{list}{Table Notes.}
\item Cols. $2-3$: Central coordinates. Cols. $4-5$: Corresponding Galactic coordinates. The full table is available online.
\end{list}
\end{table*}

\begin{figure*}
\begin{minipage}[b]{0.328\linewidth}
\includegraphics[width=\textwidth]{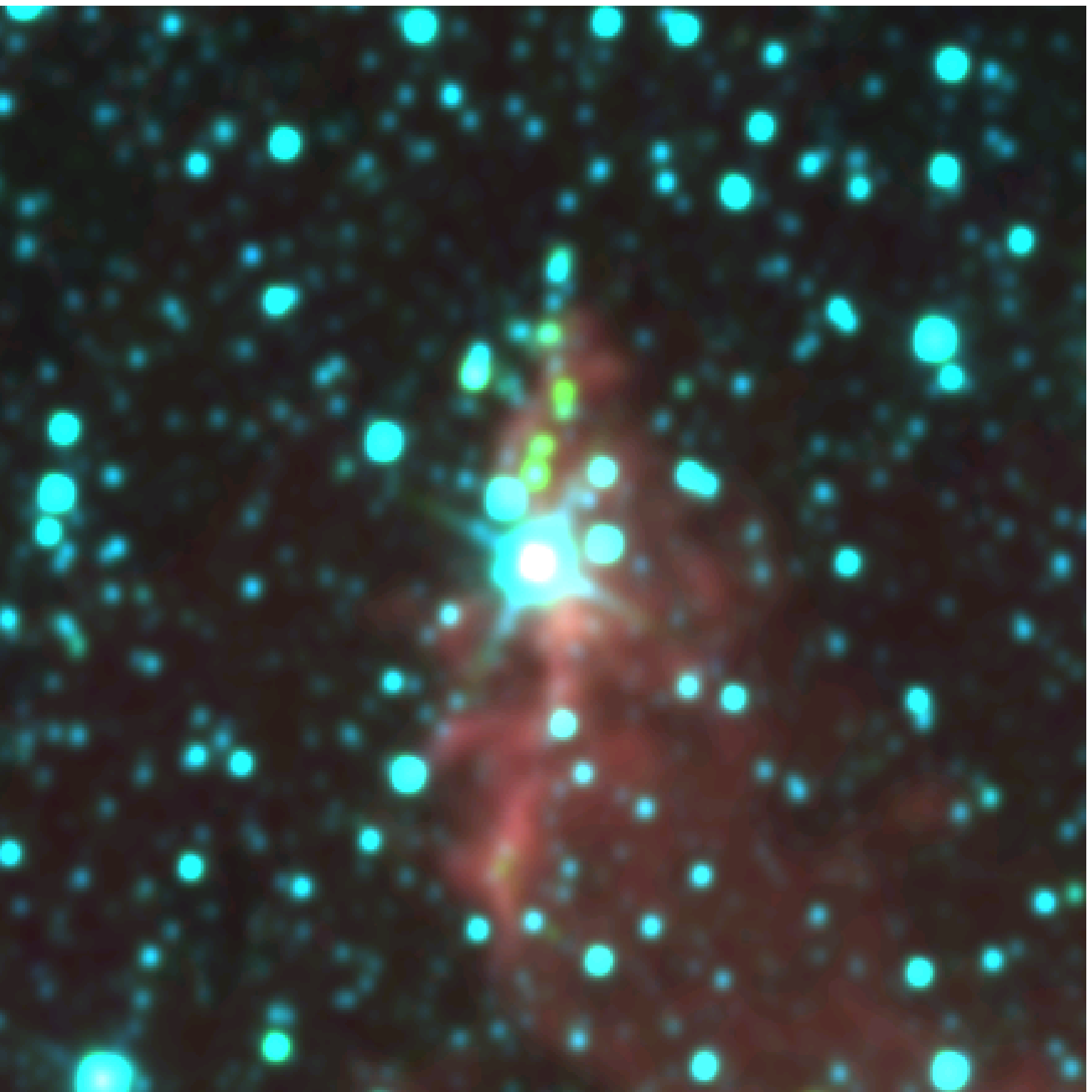}
\put(-134.0,155.0){\makebox(0.0,0.0)[5]{\fontsize{14}{14}\selectfont \color{red} C 1086}}
\end{minipage}\hfill
\vspace{0.02cm}
\begin{minipage}[b]{0.328\linewidth}
\includegraphics[width=\textwidth]{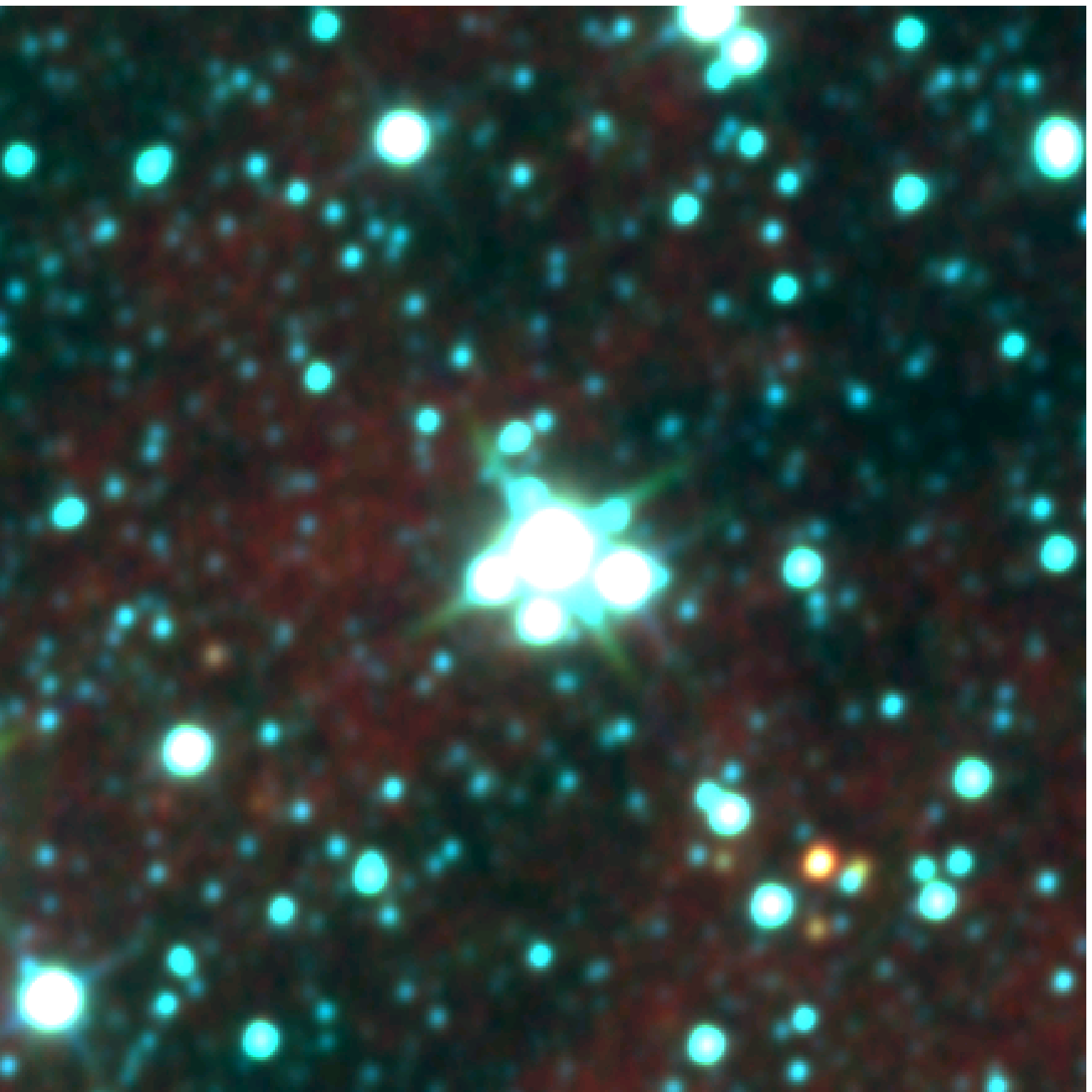}
\put(-140.0,155.0){\makebox(0.0,0.0)[5]{\fontsize{14}{14}\selectfont \color{red}C 838}}
\end{minipage}\hfill
\vspace{0.02cm}
\begin{minipage}[b]{0.328\linewidth}
\includegraphics[width=\textwidth]{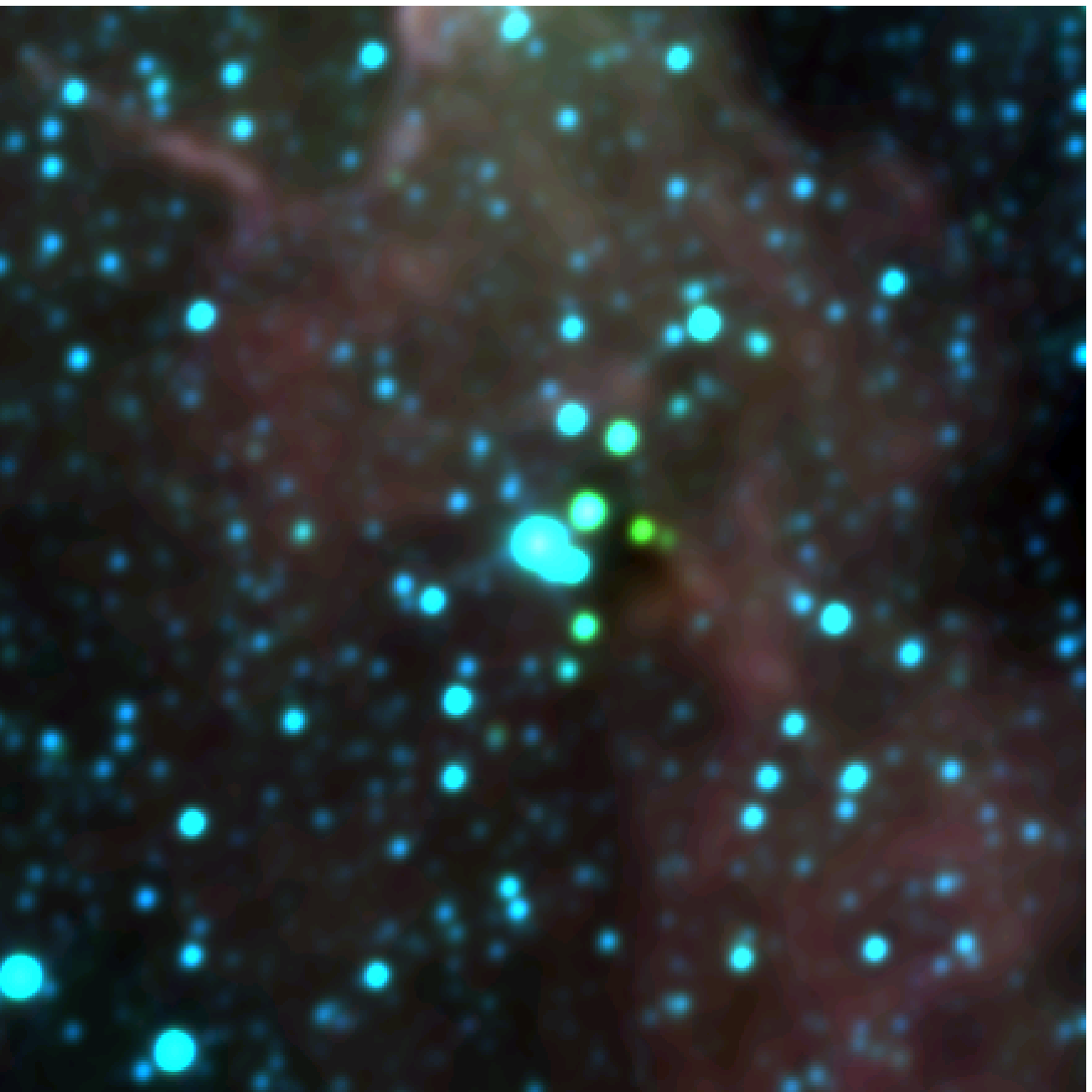}
\put(-140.0,155.0){\makebox(0.0,0.0)[5]{\fontsize{14}{14}\selectfont \color{red} C 769}}
\end{minipage}\hfill
\vspace{0.02cm}
\begin{minipage}[b]{0.328\linewidth}
\includegraphics[width=\textwidth]{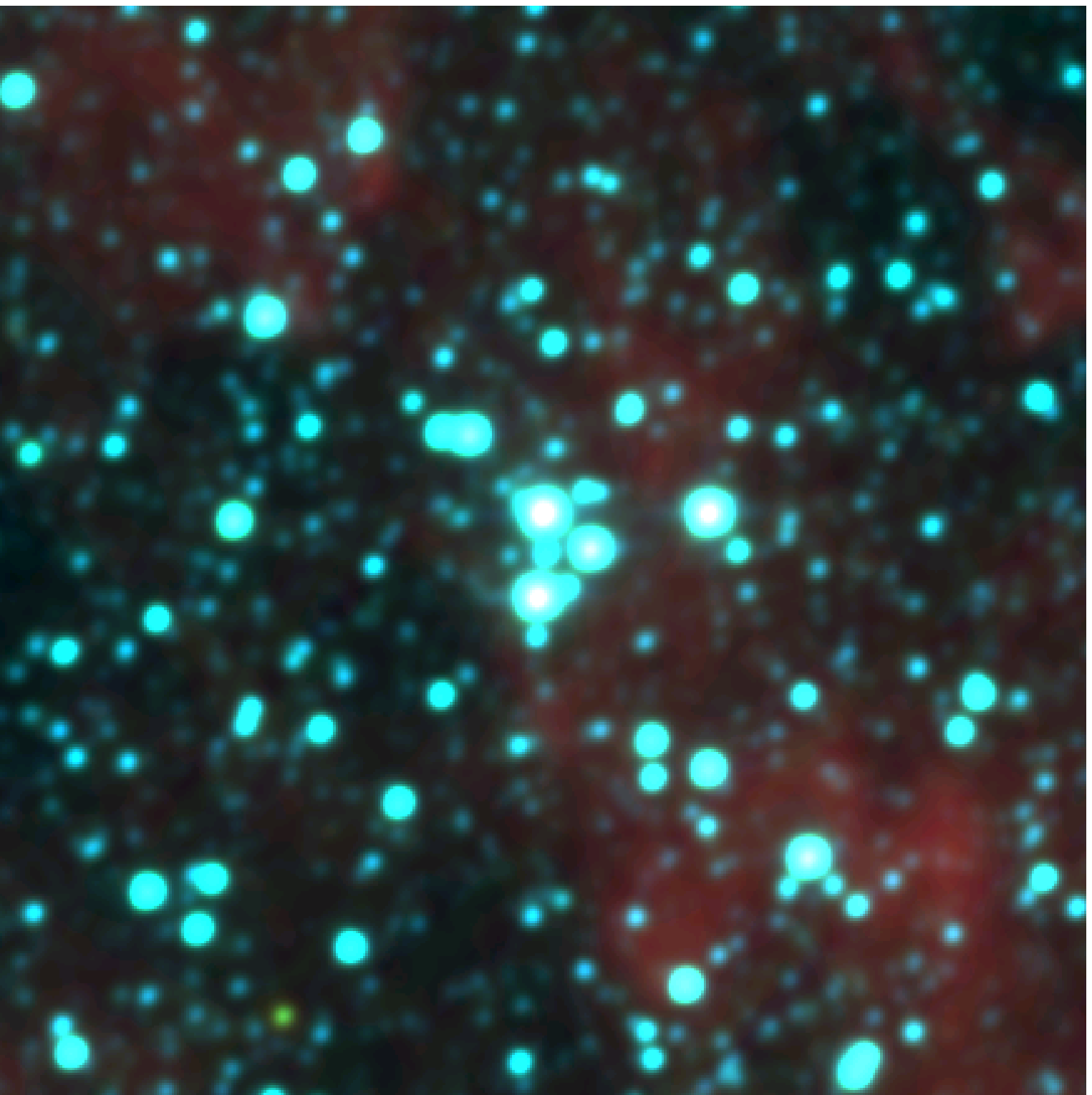}
\put(-140.0,155.0){\makebox(0.0,0.0)[5]{\fontsize{14}{14}\selectfont \color{red}C 646}}
\end{minipage}\hfill
\begin{minipage}[b]{0.328\linewidth}
\includegraphics[width=\textwidth]{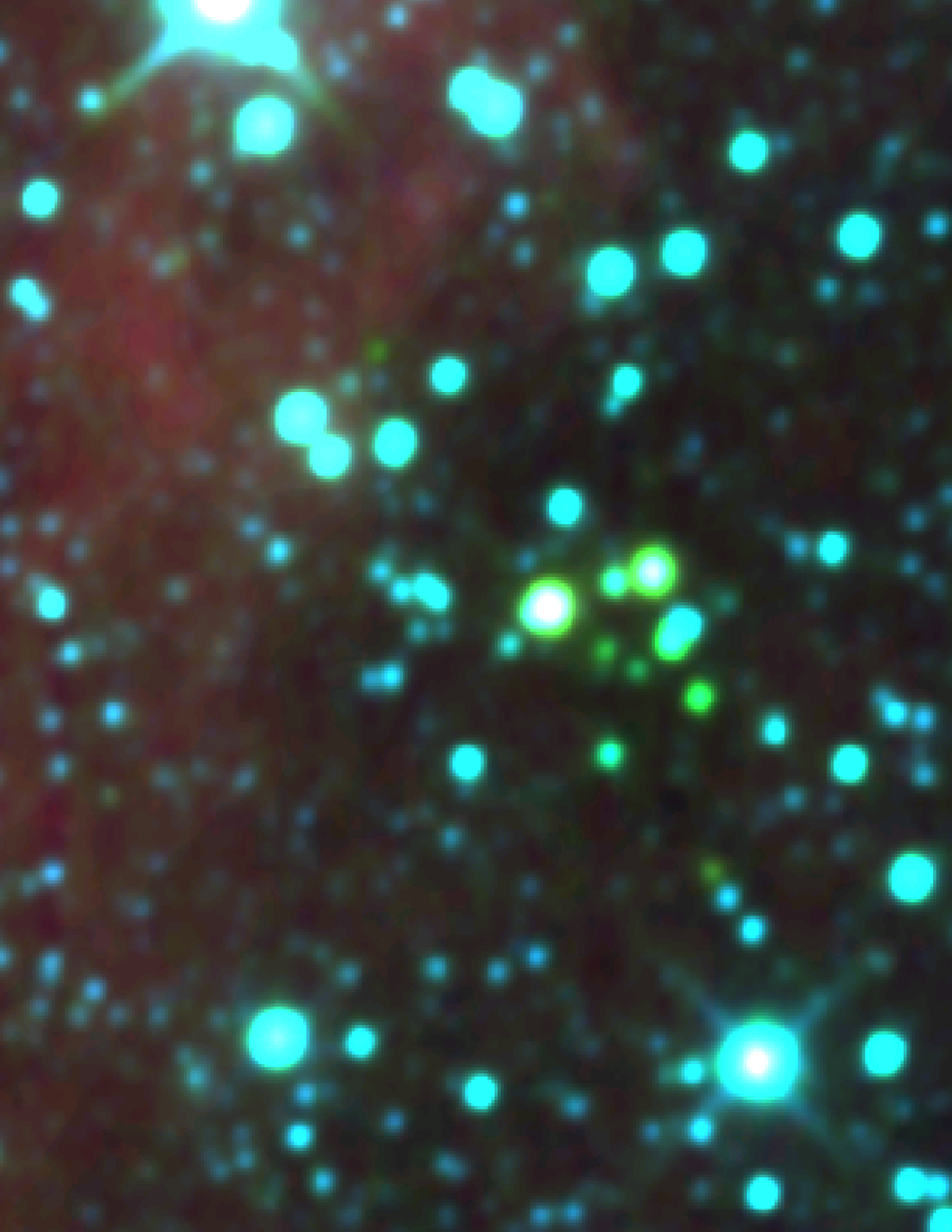}
\put(-140.0,155.0){\makebox(0.0,0.0)[5]{\fontsize{14}{14}\selectfont \color{red}C 915}}
\end{minipage}\hfill
\begin{minipage}[b]{0.328\linewidth}
\includegraphics[width=\textwidth]{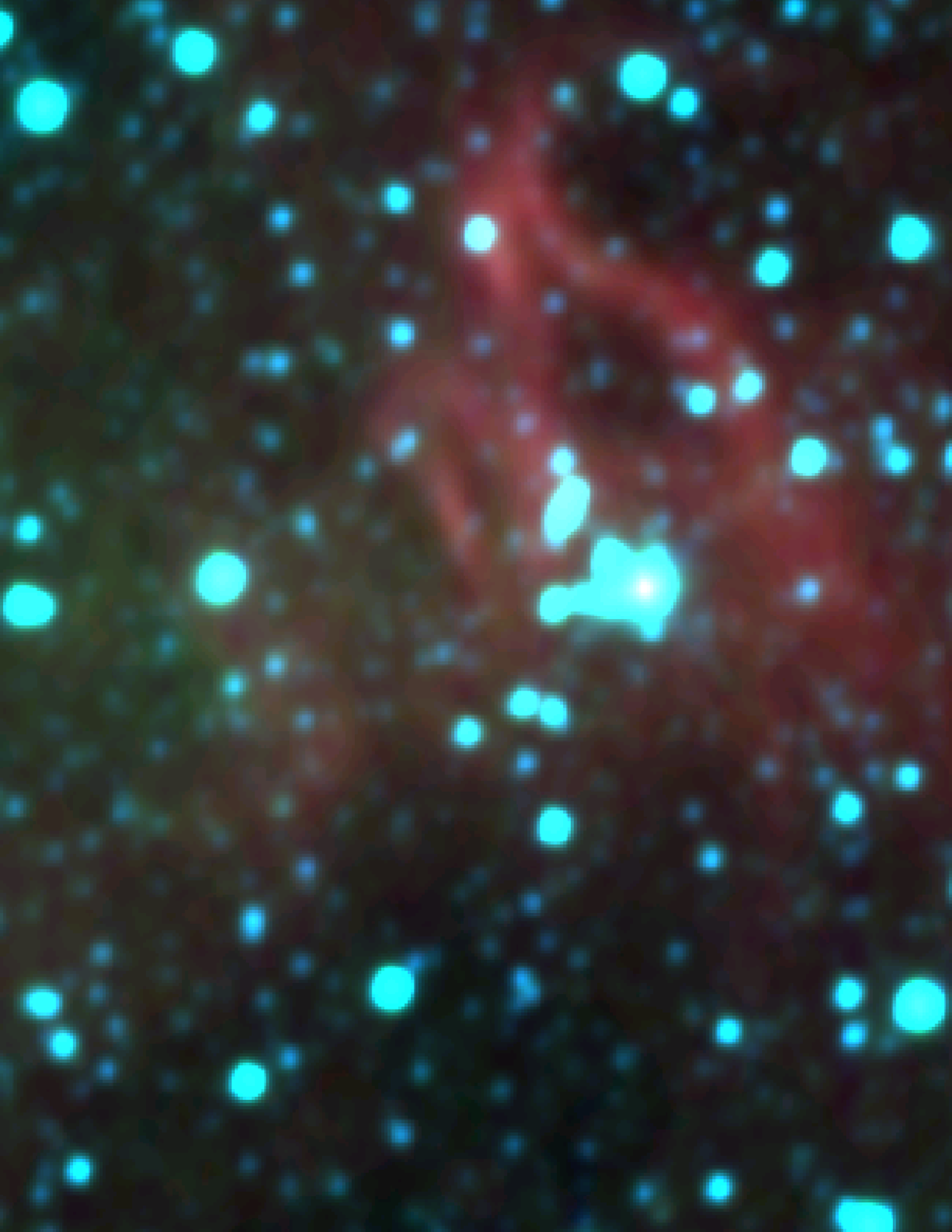}
\put(-140.0,155.0){\makebox(0.0,0.0)[5]{\fontsize{14}{14}\selectfont \color{red}C 911}}
\end{minipage}\hfill
\caption[]{Same as Fig.~\ref{f1} for the probably more evolved (less embedded) ECs C 1086, C 838, C 769, C 646, C 915, and C 911.}
\label{f2}
\end{figure*}

\begin{figure*}

\begin{minipage}[b]{0.328\linewidth}
\includegraphics[width=\textwidth]{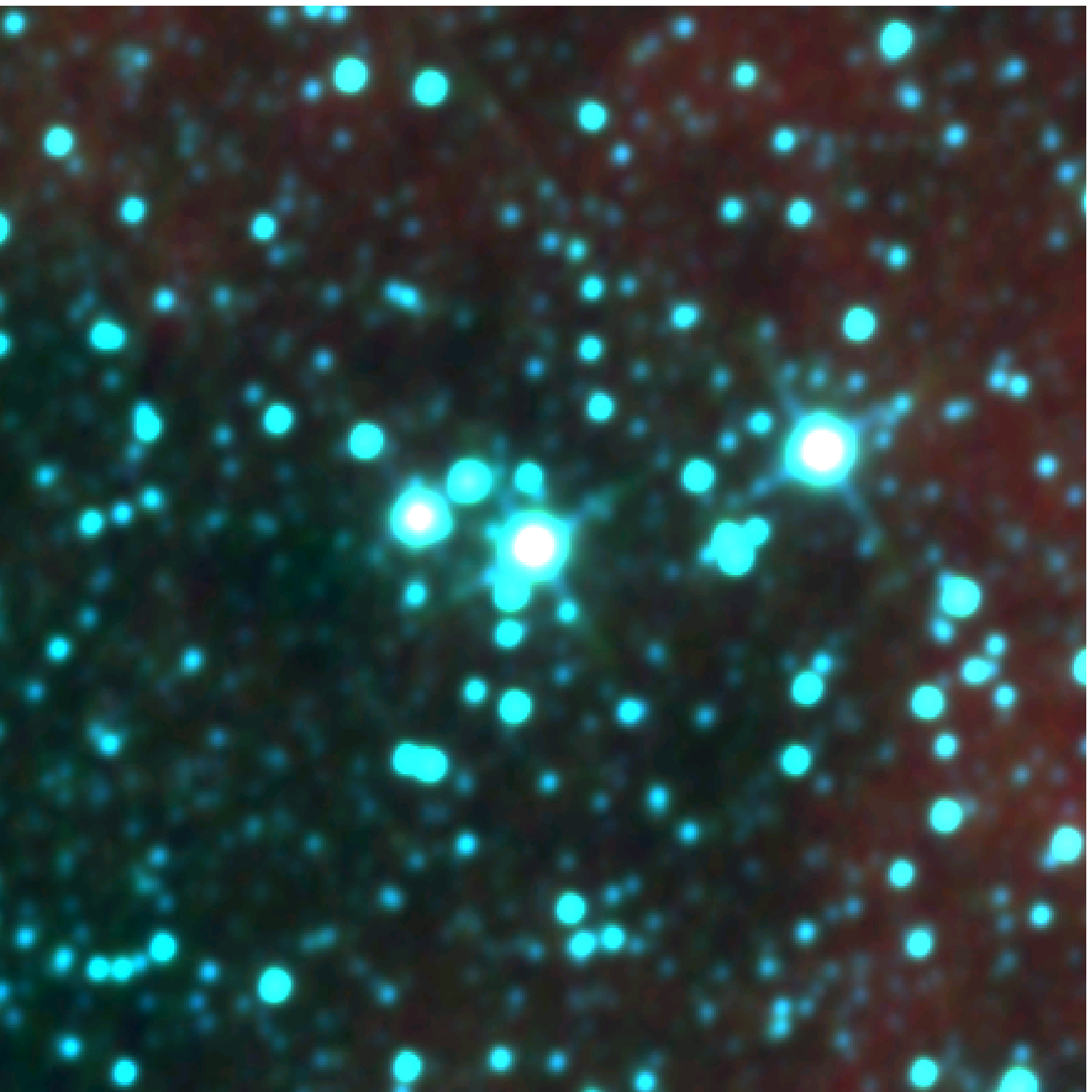}
\put(-120.0,155.0){\makebox(0.0,0.0)[5]{\fontsize{14}{14}\selectfont \color{red} C 836 (OC)}}
\end{minipage}\hfill
\vspace{0.02cm}
\begin{minipage}[b]{0.328\linewidth}
\includegraphics[width=\textwidth]{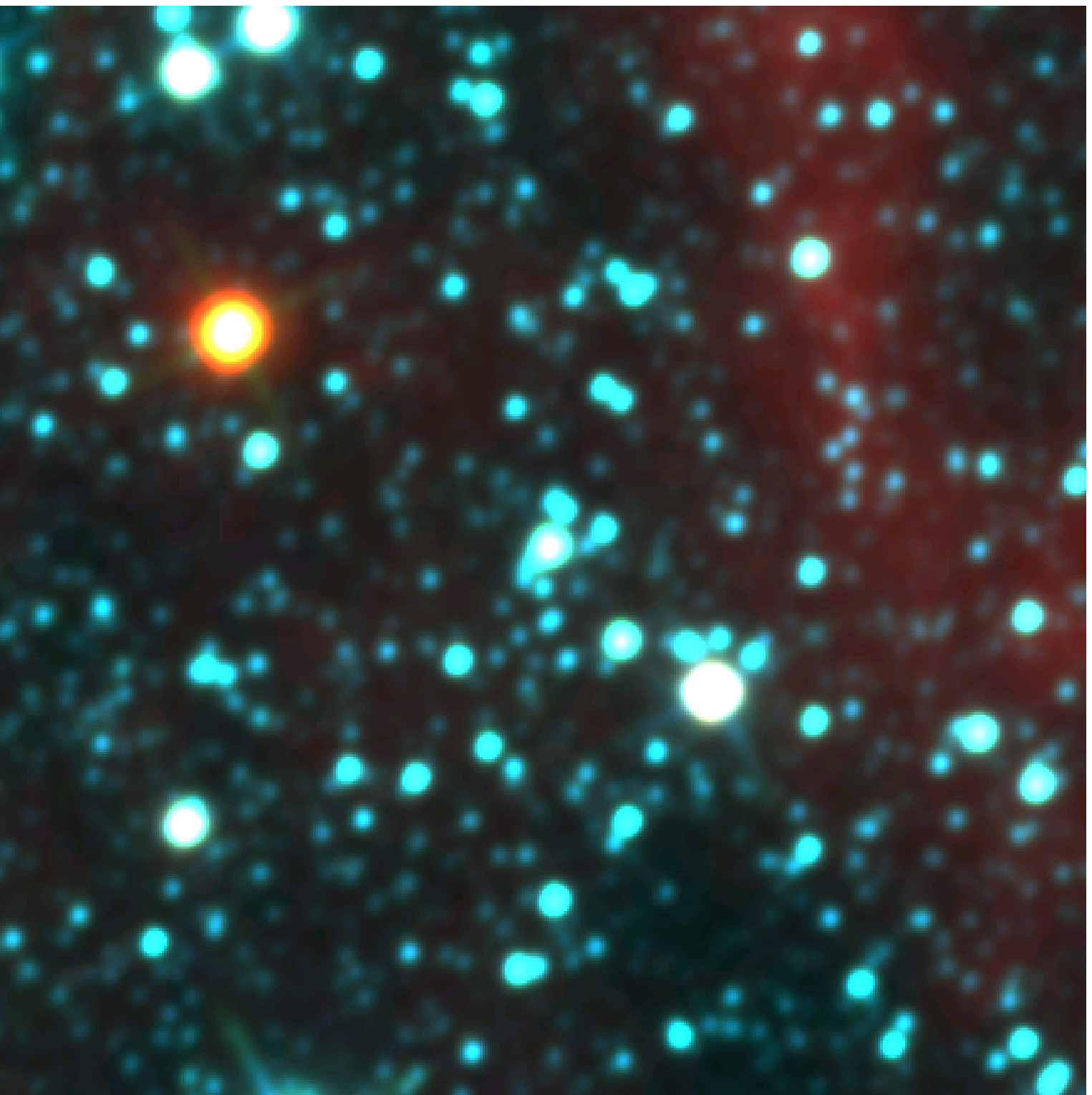}
\put(-120.0,155.0){\makebox(0.0,0.0)[5]{\fontsize{14}{14}\selectfont \color{red}C 582 (OCC)}}
\end{minipage}\hfill
\vspace{0.02cm}
\begin{minipage}[b]{0.328\linewidth}
\includegraphics[width=\textwidth]{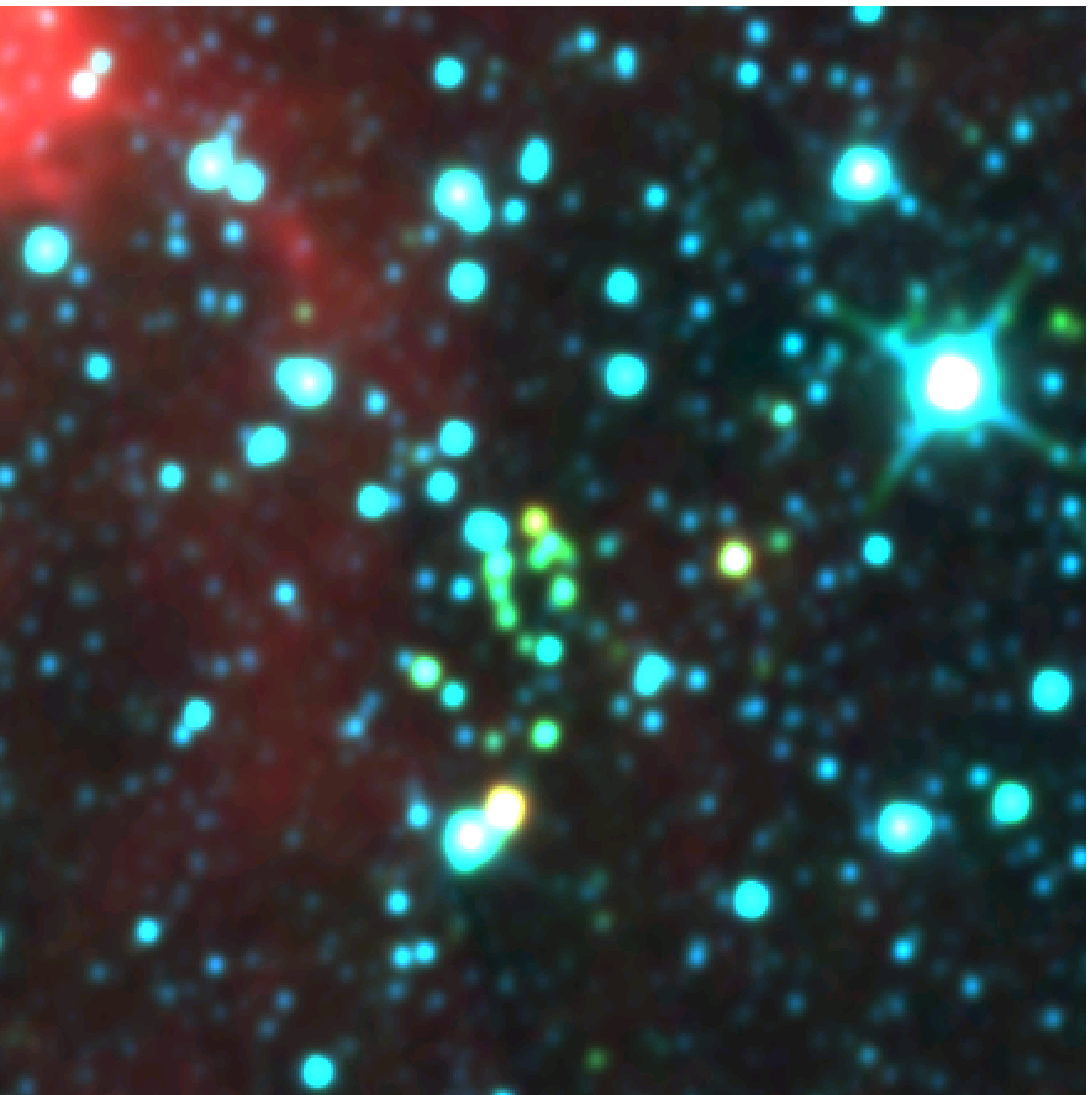}
\put(-120.0,155.0){\makebox(0.0,0.0)[5]{\fontsize{14}{14}\selectfont \color{red} C 985 (EC)}}
\end{minipage}\hfill
\vspace{0.02cm}
\begin{minipage}[b]{0.328\linewidth}
\includegraphics[width=\textwidth]{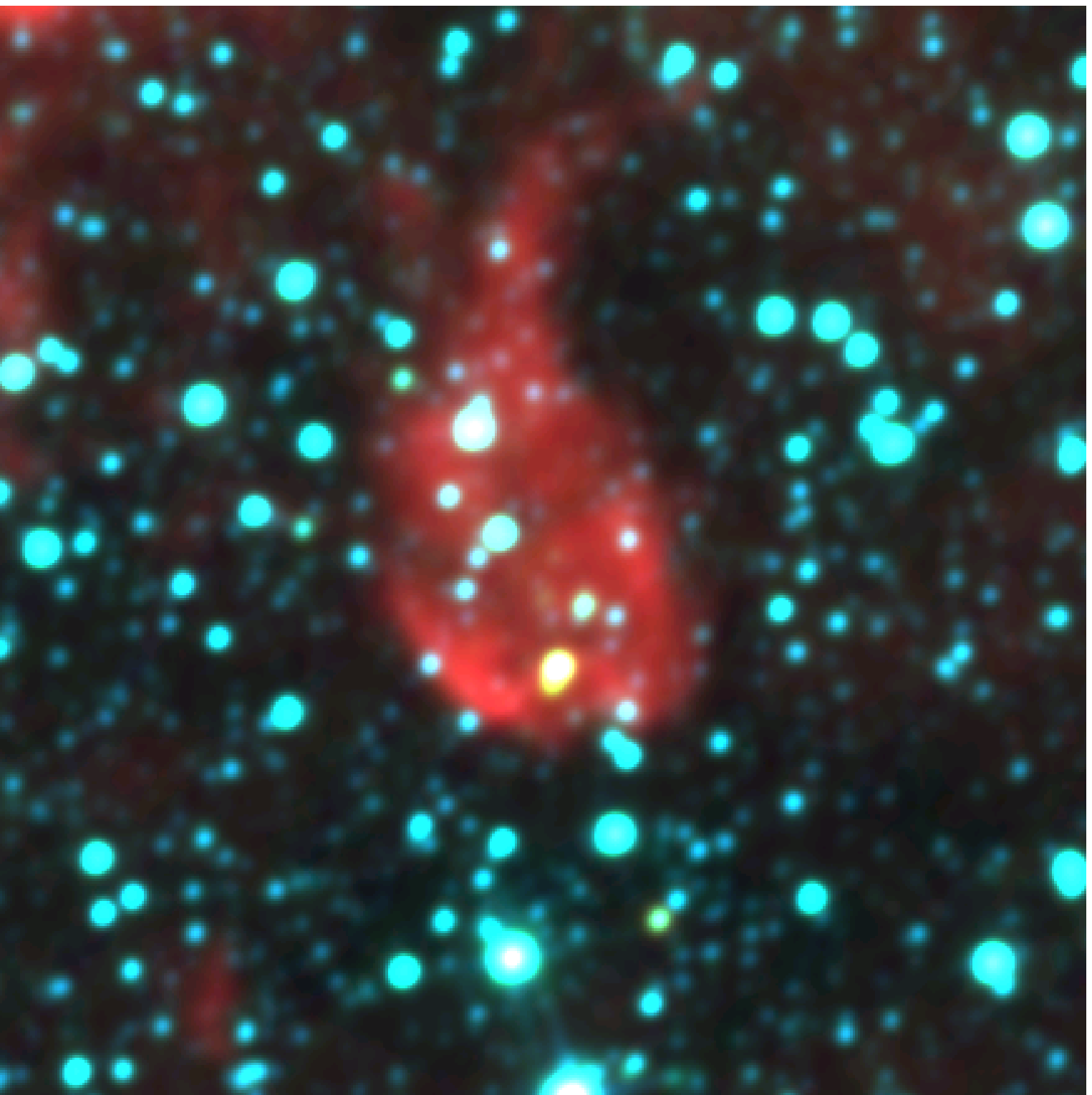}
\put(-120.0,155.0){\makebox(0.0,0.0)[5]{\fontsize{14}{14}\selectfont \color{red}C 700 (ECC)}}
\end{minipage}\hfill
\begin{minipage}[b]{0.328\linewidth}
\includegraphics[width=\textwidth]{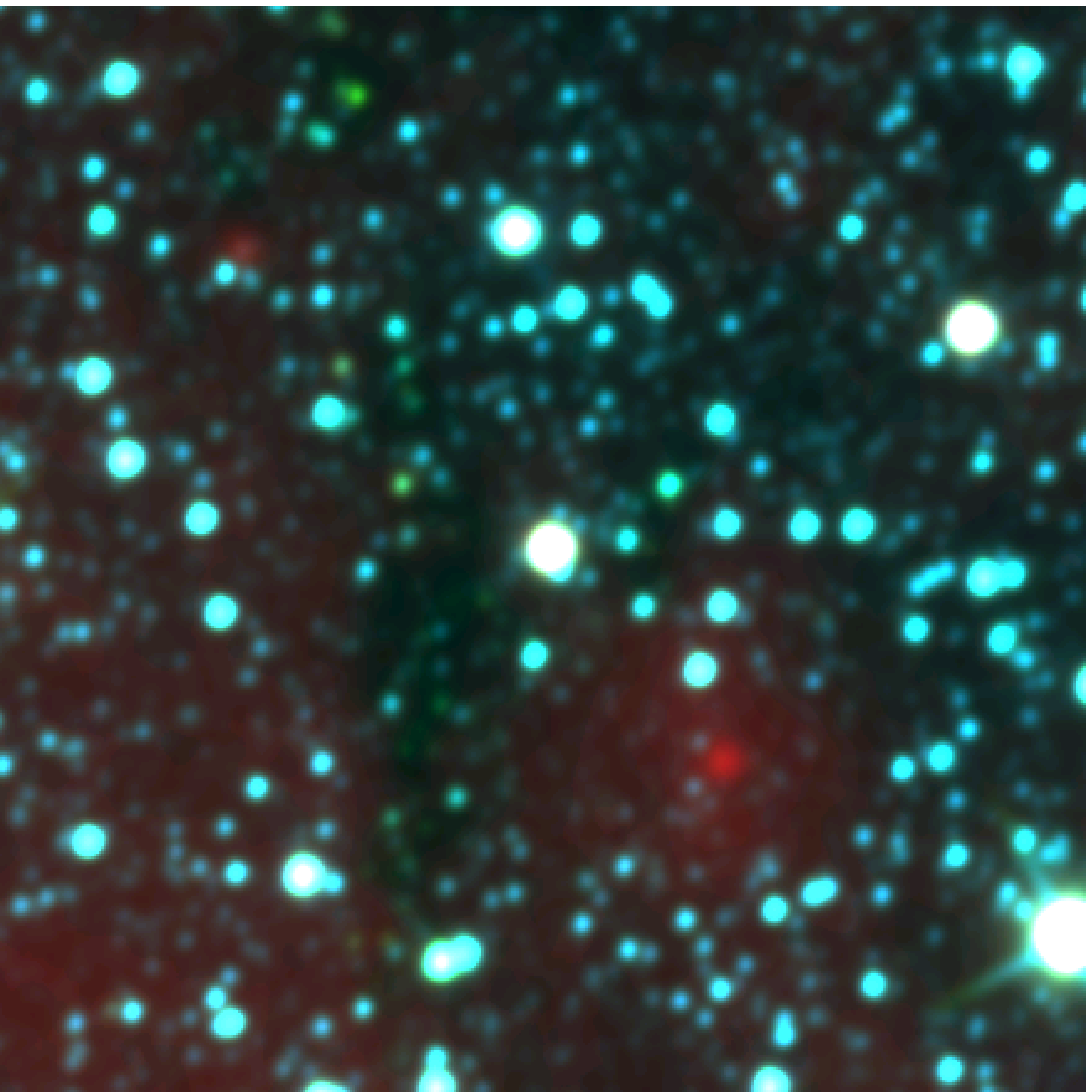}
\put(-120.0,155.0){\makebox(0.0,0.0)[5]{\fontsize{14}{14}\selectfont \color{red}C 578 (EGr)}}
\end{minipage}\hfill
\begin{minipage}[b]{0.328\linewidth}
\includegraphics[width=\textwidth]{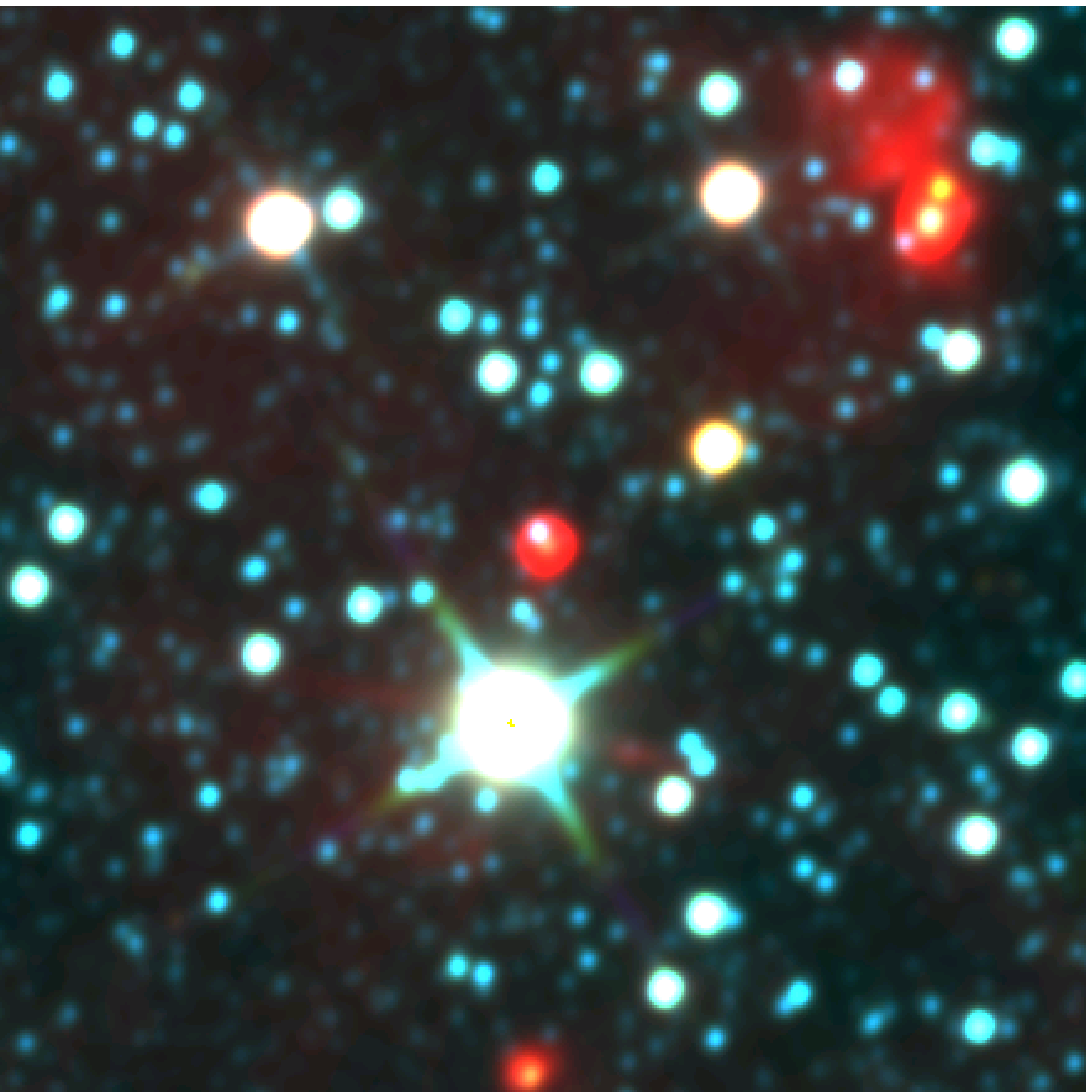}
\put(-120.0,155.0){\makebox(0.0,0.0)[5]{\fontsize{14}{14}\selectfont \color{red}C 532 (EGr)}}
\end{minipage}\hfill
\caption[]{WISE ($10'\times10'$) RGB images centred on C836 (OC), C 582 (OCC), C 985 (EC), C 700 (ECC), C 578 and C532 (EGrs): additional classifications compared to an EC.}
\label{f3}
\end{figure*}

\begin{figure*}

\begin{minipage}[c][12cm][t]{.49\textwidth}
  \centering
  \includegraphics[width=11.6cm,height=11.6cm]{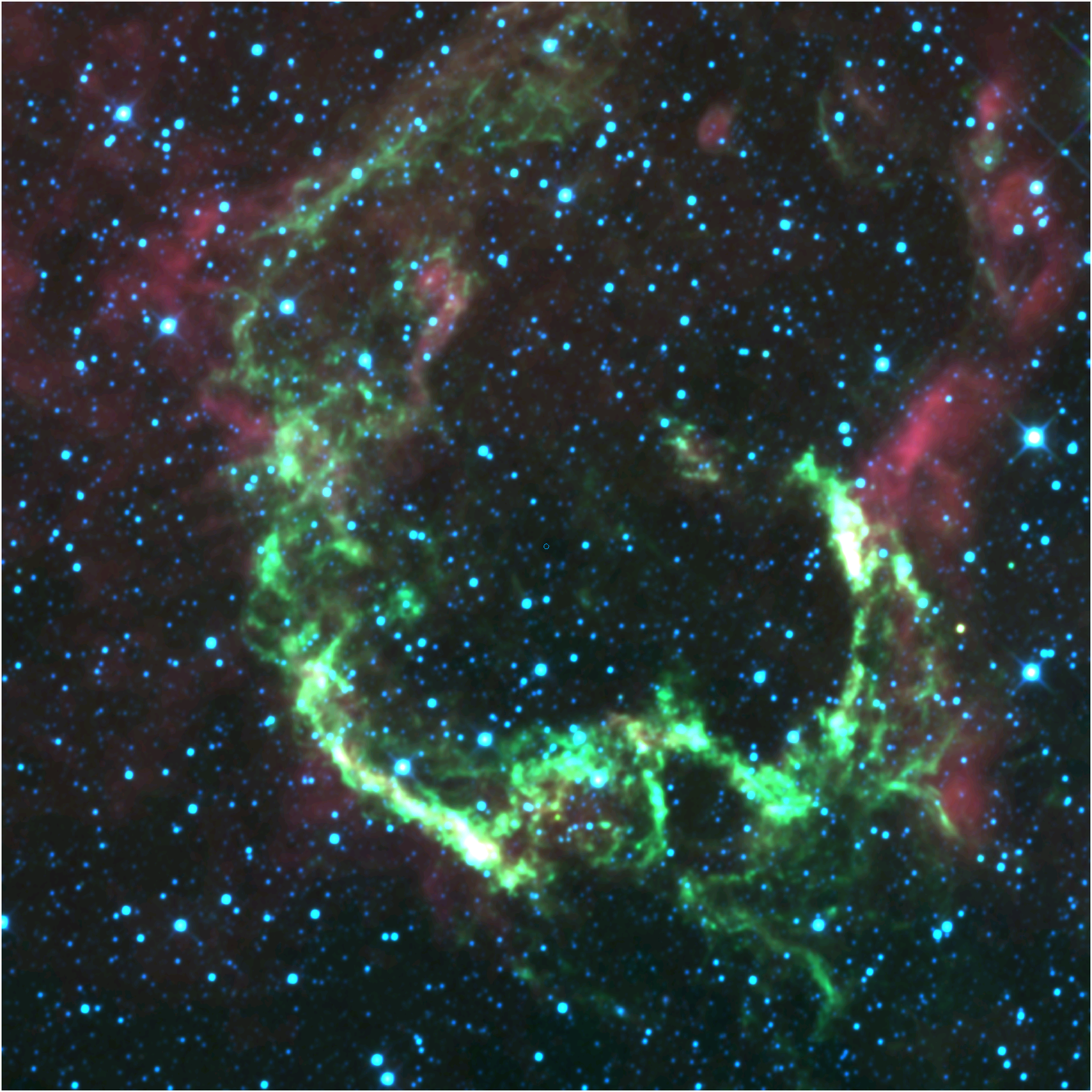}
\put(-95.0,165.0){\makebox(0.0,0.0)[5]{\fontsize{10}{10}\selectfont \color{red}C 935}}
\put(-100.0,75.0){\makebox(0.0,0.0)[5]{\fontsize{10}{10}\selectfont \color{red}C 936}}
\put(-265.0,125.0){\makebox(0.0,0.0)[5]{\fontsize{10}{10}\selectfont \color{red}C 937}}
\put(-205.0,70.0){\makebox(0.0,0.0)[5]{\fontsize{10}{10}\selectfont \color{red}C 938}}
\put(-165.0,118.0){\makebox(0.0,0.0)[5]{\fontsize{10}{10}\selectfont \color{red}FSR 891}}
\put(-152.0,107.0){\makebox(0.0,0.0)[5]{\fontsize{10}{10}\selectfont \color{red}\circle{6}}}
\put(-217.0,221.0){\makebox(0.0,0.0)[5]{\fontsize{10}{10}\selectfont \color{red}\circle{6}}}
\put(-205.5,98.5){\makebox(0.0,0.0)[5]{\fontsize{10}{10}\selectfont \color{red}\circle{6}}}
\put(-276.5,231.3){\makebox(0.0,0.0)[5]{\fontsize{10}{10}\selectfont \color{red}\circle{6}}}
\put(-43.5,46.8){\makebox(0.0,0.0)[5]{\fontsize{10}{10}\selectfont \color{red}\circle{6}}}
\end{minipage}
\vspace{0.03cm}
\begin{minipage}[c][12cm][t]{.49\textwidth}
  \raggedleft
  \includegraphics[width=5.76cm,height=5.76cm]{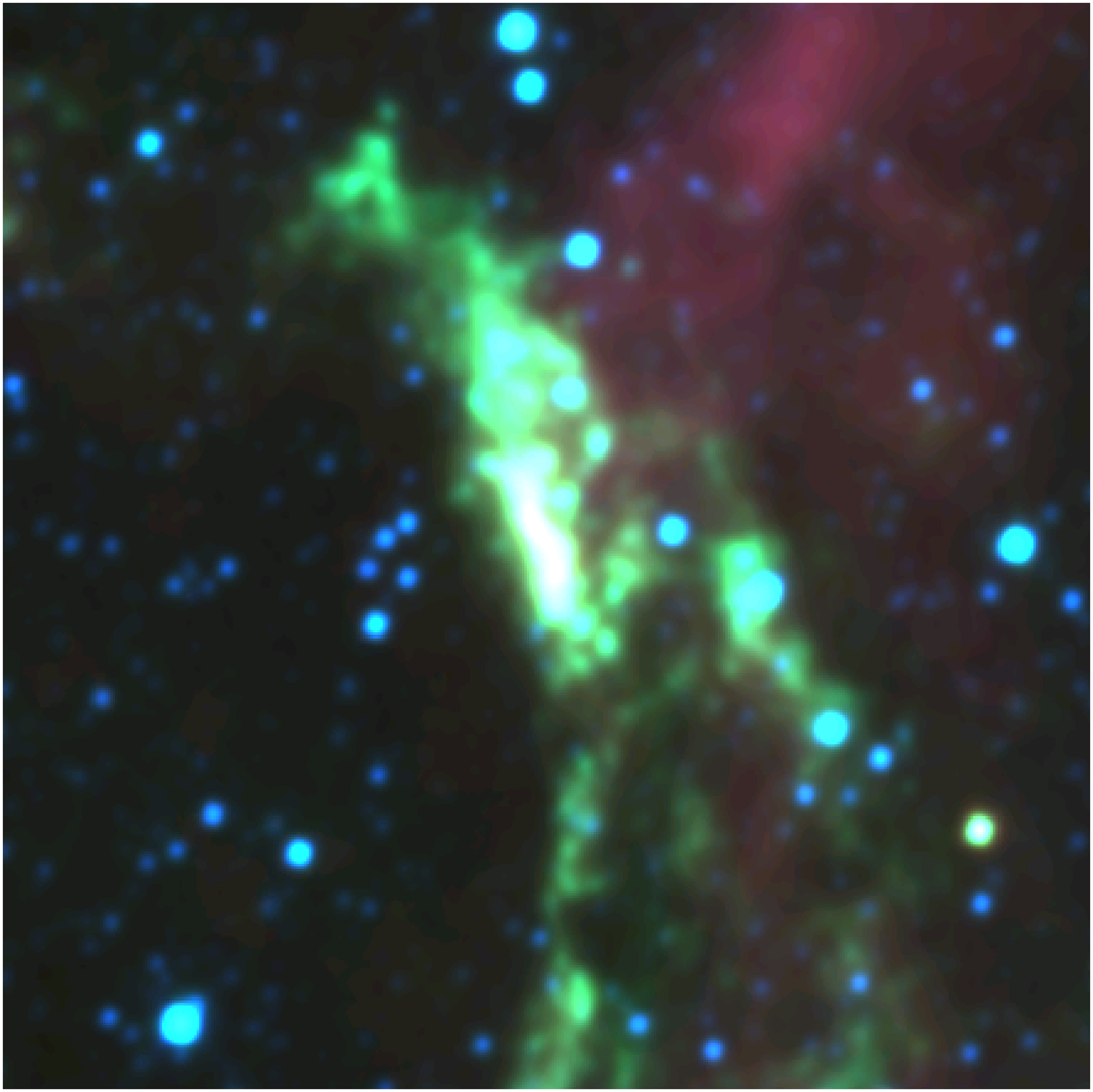}
\put(-138.0,155.0){\makebox(0.0,0.0)[5]{\fontsize{10}{10}\selectfont \color{red}C 935}}
\hspace{0.5cm}
  \includegraphics[width=5.76cm,height=5.76cm]{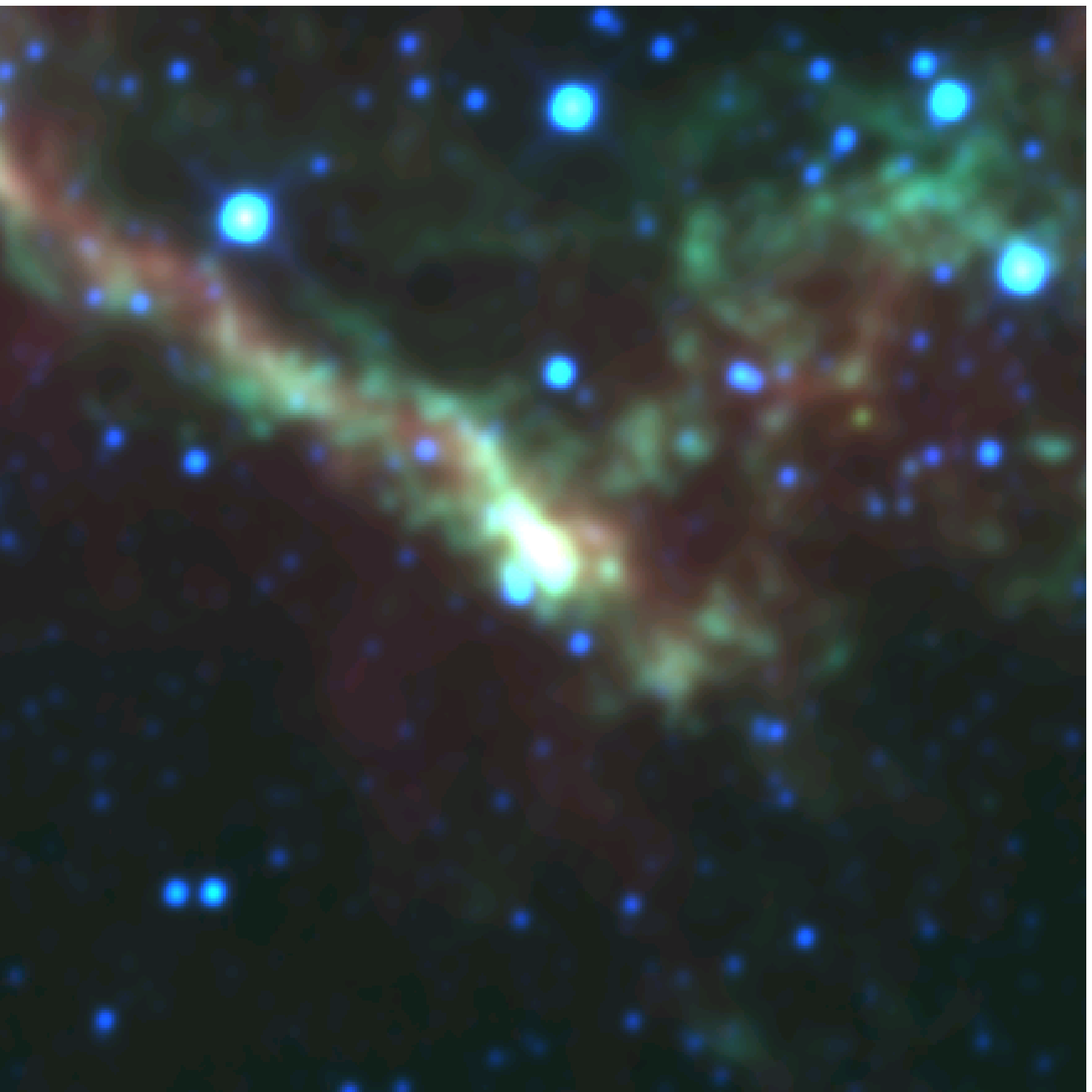}
\put(-138.0,155.0){\makebox(0.0,0.0)[5]{\fontsize{10}{10}\selectfont \color{red}C 938}}
\put(-122.0,131.7){\makebox(0.0,0.0)[5]{\fontsize{10}{10}\selectfont \color{red}\circle{10}}}
\end{minipage}
\caption{Shell with C 935, C 936, C 937, and C 938: formation  of a cluster aggregate.
Top right: blowup of C 935. Bottom right: blowup of C 938. Circles indicate OB stars from SIMBAD.}
\label{f4}
\end{figure*}

\begin{figure*}

\begin{minipage}[c][12cm][t]{.49\textwidth}
  \centering
  \includegraphics[width=11.6cm,height=11.6cm]{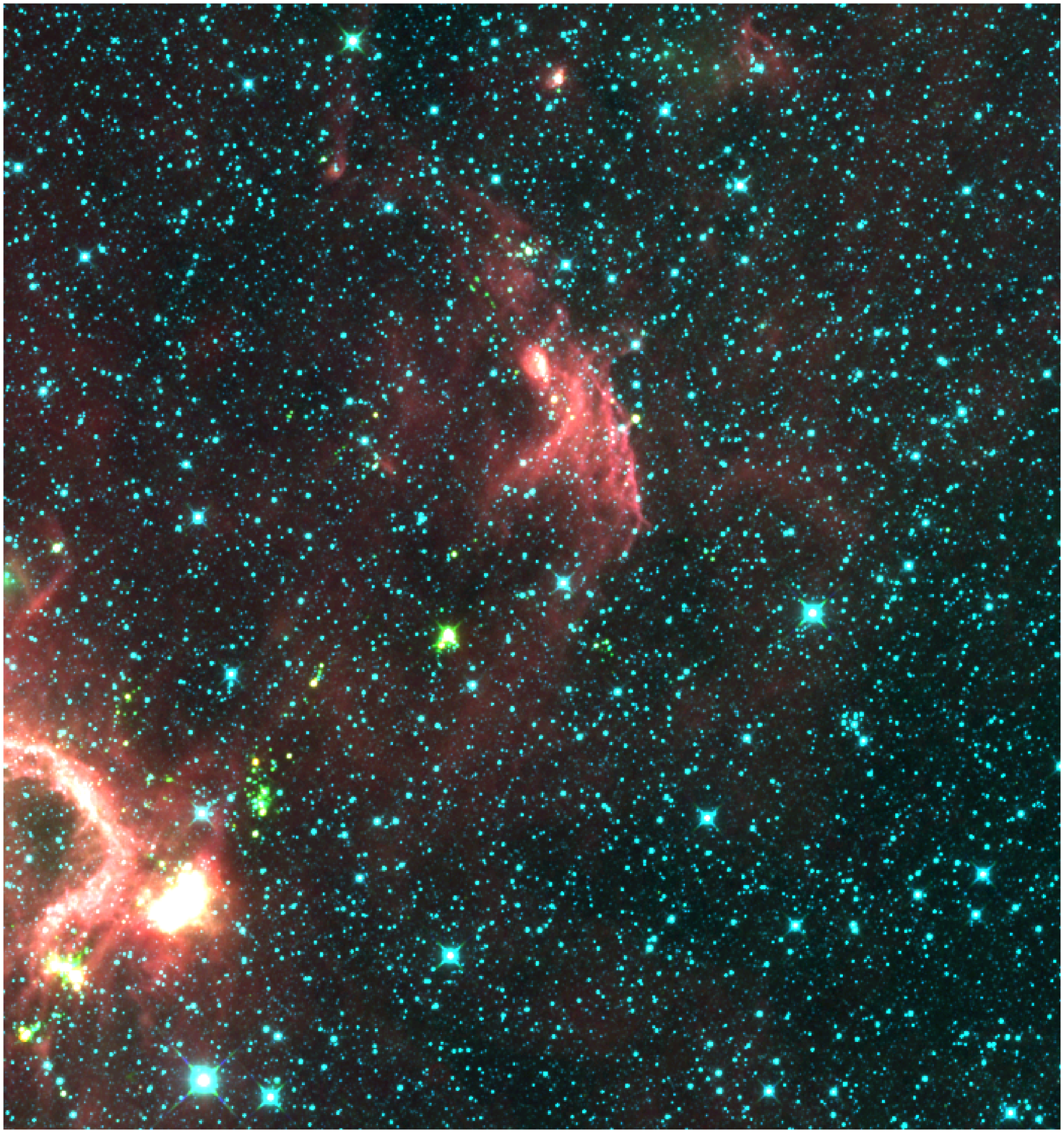}
\put(-205.0,170.0){\makebox(0.0,0.0)[5]{\fontsize{10}{10}\selectfont \color{yellow}C 916}}
\put(-250.0,62.0){\makebox(0.0,0.0)[5]{\fontsize{10}{10}\selectfont \color{yellow}C 925}}
\put(-142.0,135.0){\makebox(0.0,0.0)[5]{\fontsize{10}{10}\selectfont \color{yellow}C 915}}
\put(-235.0,200.0){\makebox(0.0,0.0)[5]{\fontsize{10}{10}\selectfont \color{yellow}C 914}}
\put(-257.0,45.0){\makebox(0.0,0.0)[5]{\fontsize{10}{10}\selectfont \color{yellow}Sh2-233 SE Cluster}}
\put(-265.0,28.0){\makebox(0.0,0.0)[5]{\fontsize{10}{10}\selectfont \color{yellow}G173.58+2.45 Cluster}}
\put(-235.0,125.0){\makebox(0.0,0.0)[5]{\fontsize{10}{10}\selectfont \color{yellow}C 919}}
\put(-295.0,115.0){\makebox(0.0,0.0)[5]{\fontsize{10}{10}\selectfont \color{yellow}C 921}}
\put(-190.0,132.0){\makebox(0.0,0.0)[5]{\fontsize{10}{10}\selectfont \color{yellow}SUH 124}}
\put(-220.0,95.0){\makebox(0.0,0.0)[5]{\fontsize{10}{10}\selectfont \color{yellow}SUH 162}}
\end{minipage}%
\vspace{0.03cm}
\begin{minipage}[c][12cm][t]{.49\textwidth}
  \raggedleft
\includegraphics[width=5.76cm,height=5.76cm]{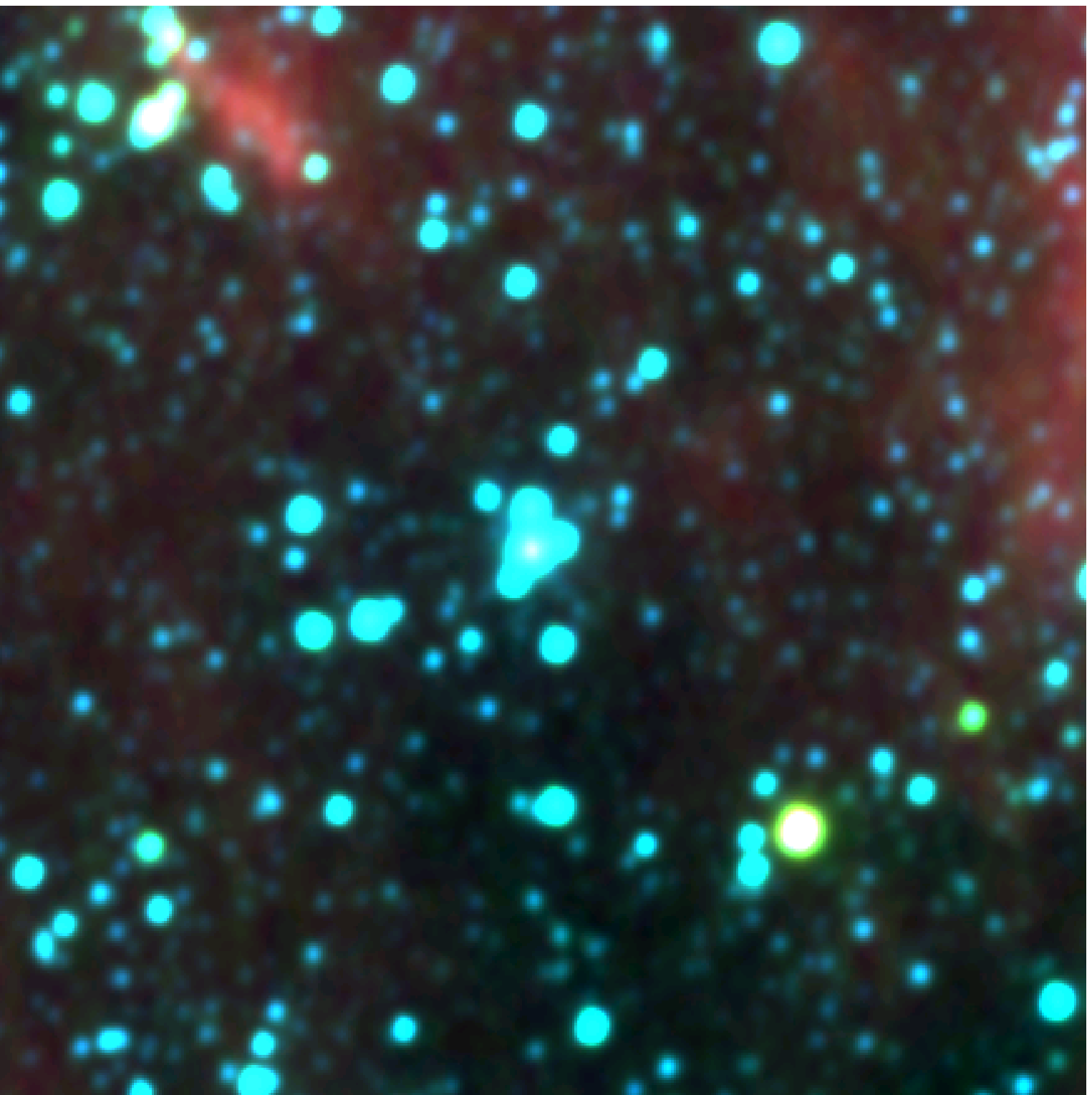}
\put(-22.0,155.0){\makebox(0.0,0.0)[5]{\fontsize{10}{10}\selectfont \color{yellow}C 916}}
\hspace{0.8cm}
\includegraphics[width=5.76cm,height=5.76cm]{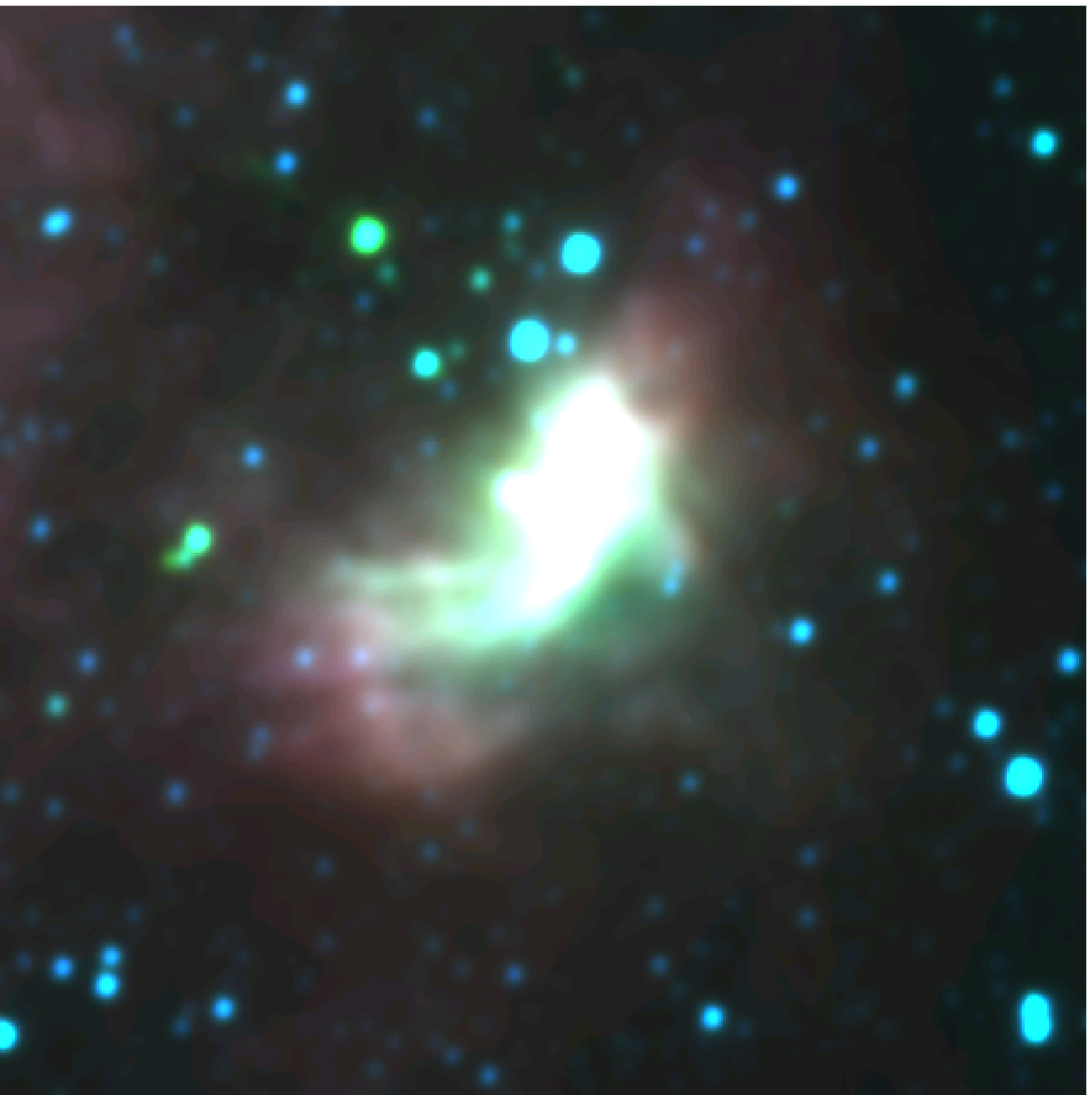}
\put(-22.0,155.0){\makebox(0.0,0.0)[5]{\fontsize{10}{10}\selectfont \color{yellow}C 925}}
\end{minipage}
\caption{Left panel ($2^{\circ}\times2^{\circ}$): dust cloud forming Cluster Aggregate: C 914, C 915, C 916, C 919, C 921, C 925, Sh2-233 SE Cluster, $G173+2.45$ Cluster, SUH 124 and SUH 162.
Top right: blowup of C 916 ($10'\times10'$), Bottom right: blowup of C 925 ($10'\times10'$).}
\label{f5}
\end{figure*}

The present work deals with new discoveries of star clusters, mostly ECs with WISE and analyses of their properties, following  the series of recent papers \citet{Camargo15a, Camargo15b, Camargo15c}, hereafter Papers I, II and III, respectively.

The paper is organized as follows. In Sect. \ref{sec:2} we present the search procedure and the newly found clusters. In Sect. \ref{sec:3} we analyse a subsample of the discovered clusters using  2MASS photometry. We also describe the methods employed in the cluster analyses. They  employ Colour-magnitude Diagrams (CMDs) and Radial Density Profiles (RDP). Sect. \ref{sec:4} is dedicated to the discussion of the results. Finally, in Sect. \ref{sec:5} we provide the concluding remarks.

\section{Present work discoveries}
\label{sec:2}

We recently reported the discovery of 446 star clusters, stellar groups and candidates, mostly projected on the Galactic disk. They were designated Camargo 1 (C 1)  to C 446, in a series of analyses: 
\textit{(i)} In Paper I we presented the first  large list: C 1 to C 437.
\textit{(ii)} In Paper II we analysed C 438 and C 439, two rare high Galactic latitude embedded clusters that were formed in the halo.
\textit{(iii)} In Paper III we provided the ECs C 440 to C 446, in a study of Spiral Arm tracers.

The present paper discoveries  of star clusters and alike are designated  C 447 to C 1098, thus amounting to 652 new entries (Table~\ref{tab1}).

Our initial sample in the present study consisted of 862 objects that we detected on the WISE multi-band Atlas, along the four Galactic quadrants, in general for $|b|<30^{\circ}$. The subsequent analyses indicated that 212 had already been reported in the literature, or were discarded owing to miscellaneous reasons for a few of them, like probable planetary nebulae or spurious objects when inspected in detail. Thus, $75\%$ of the initial sample were considered to be new star clusters, EGrs  or candidates, and were included in the present catalogue. We list in Table~\ref{tab1} 38 entries that are shown as examples or discusssed in more detail throughout the present paper. The complete Table~\ref{tab1} will be available in the online version.

In Figs.~\ref{f1} to \ref{f5} are shown examples of the discovered clusters and alike. All objects
were found by means of visual inspections on WISE \citep{Wright10}, utilizing  image services  at NASA/IPAC or Aladin. The W1 ($3.4{\mu}$m) and W2 ($4.6{\mu}$m) bands are more sensitive to the stellar and PMS content, while in W3 ($12{\mu}$m) and W4 ($22{\mu}$m) extended structures arise mostly from dust emission.

We started out by  looking for dust emission nebulae on WISE, followed by a search for stellar overdensities within them. We cross-correlated our objects with previous catalogues or lists \citep[][and Paper I]{Acker92, Bica03a, Bica03b, Dutra03, Bica05, Kharchenko05a, Kharchenko05b, Mercer05, Kronberger06, Froebrich07, Koposov08, Glushkova10, Borissova11, Majaess13}.
Searches for new star clusters often employ automatic softwares that identify stellar overdensities \citep[e.g.][]{Froebrich07}. However, heavily dust-obscured ECs may be missed by an automatic algorithm, especially with 2MASS, since a significant fraction of the stellar content may not be accessible. As a consequence, stellar detections in such deeply embedded clusters may be  underestimated with respect to the surrounding field.
Following Paper I we tentatively classified the discovered objects in open clusters (OCs), open cluster candidates (OCCs), embedded clusters (ECs), embedded cluster candidates (ECCs), and embedded stellar groups (EGrs). This classification is based on the stellar/dust density estimates on the WISE bands (W1 to W4) and RGB (Red Green and Blue) images (see Paper I). Similar criteria were used in previous works \citep{Hodapp94, Bica03a, Dutra03}. 

In Figs.~\ref{f1} and \ref{f2} we show a representative sample of  newly found objects to exemplify the EC classification.
In Fig.~\ref{f1} the stellar content is more embedded than in Fig.~\ref{f2}, which suggests evolutionary effects.
Fig.~\ref{f1}  shows six compact relatively isolated ECs. These objects are small dusty and the stellar core is deeply embedded in their natal nebulae. For instance, C 941, is a very small and poor EC. The structures of ECs in Fig.~\ref{f2} are  dominated by compact cores of stars. In Fig.~\ref{f3} we show additional objects (Table~\ref{tab1}) to clarify the classification, an OC and an OCC without dust emission, an EC with a prominent core, a poor ECC within a dust nebula, and two EGrs that are less dense than the EC.  Figs.~\ref{f4} and \ref{f5} are dedicated to illustrate EC aggregates \citep[][Paper III]{Camargo11, Camargo12}. 
In the large panel of Fig.~\ref{f5} we show C 914, C 915, C 916, C 919, C 921, and C 925, together with Sh2-233 SE Cluster, $G173+2.45$ Cluster, SUH 124 and SUH 162 \citep{Solin12}.
These ECs are located  in the Eastern half of a large star forming complex, where yet more ECs exist. Likewise, in Fig.~\ref{f4}, we appear to be  witnessing  the birth of a Cluster Aggregate, however with less members. The emission dust  shell suggests  that these ECs are  coeval.
 In Fig.~\ref{f6} we show three interesting closely projected  star forming clumps, which may be an example of multiple cluster formation or just a snapshot view of massive cluster evolution. Such environments differ from Cluster Aggregates in the sense that their compact nature would favour more frequent cluster interactions, such as mergers.

Fig.~\ref{f7} shows the angular distribution of the present sample, as compared to the other large WISE cluster surveys \citep[][Paper I]{Majaess13}. A few high Galactic objects are beyond $|b|=30^{\circ}$ (Table~\ref{tab1}), and will be discussed elsewhere.

\section{Analisys of a representative sample}
\label{sec:3}

We use 2MASS\footnote{The Two Micron All Sky Survey.} photometry \citep{Skrutskie06} in the $J$, $H$ and $K_{s}$ bands to analyse the nature of a representative cluster sample. Stars are extracted in circular regions centred in the coordinates of each cluster candidate.

In view of building the intrinsic CMD the observed photometry is submitted to a field-star decontamination procedure, which is described in detail in \citet{Bonatto07a, Bonatto07b} and \citet{Bica08}. The fundamental parameters are derived by fitting PARSEC isochrones \citep{Bressan12} to the $J\times(J-K_s)$ decontaminated CMDs. The cluster structures are analysed via RDP.

\begin{figure}
\resizebox{\hsize}{!}{\includegraphics{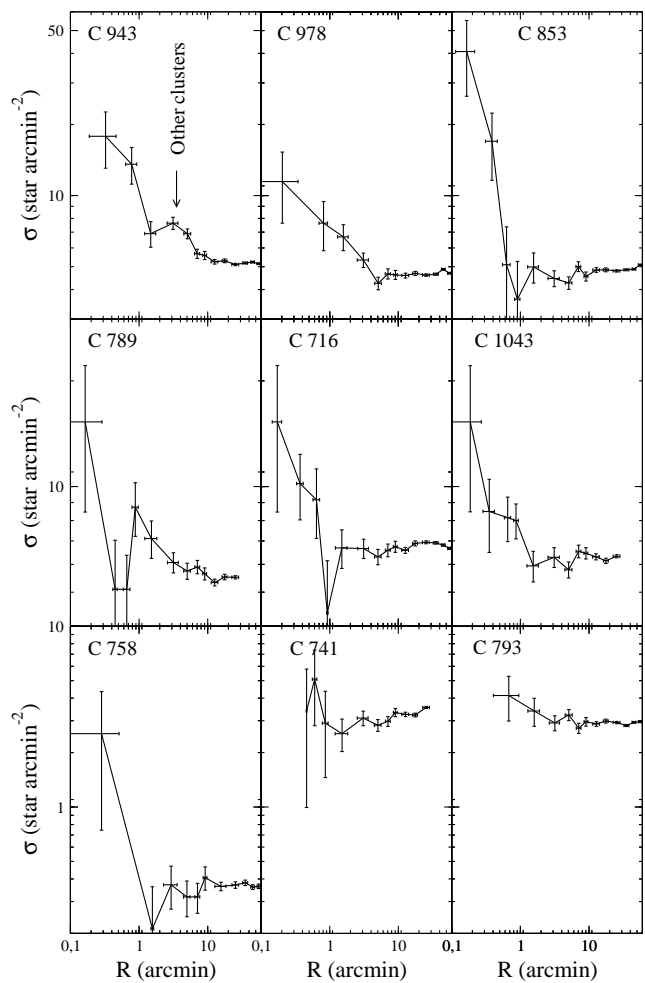}}
\caption[]{Radial Density Profiles (RDPs) for embedded clusters in Table~\ref{tab2}.}
\label{f11}
\end{figure}

Fig.~\ref{f8} shows WISE images of six ECs from the representative sample in Table~\ref{tab2} while in Figs.~\ref{f9}, \ref{f10}, and \ref{f11} are shown their decontaminated CMDs and RDPs. In Table~\ref{tab2} we show the derived cluster parameters from the CMD. 
The clusters are younger than 5 Myr, and some are as young as 1-2 Myr. Distances
are in the range $d_{\odot}=2.8-6.0$ kpc. The RDPs are typical of  well-studied embedded clusters \citep[e.g.][]{Bonatto09, Bonatto11}, showing a central peak and at times radial dips possibly caused by dust absorption or crowding.

\section{Discussion}
\label{sec:4}
\subsection{Relatively Isolated ECs}

The ISM consists of a hierarchical structure of gas and dust, from giant molecular cloud complexes to small cores. Since star clusters emerge from these structures we can expect some similarity mainly in the embedded phase. These objects may span a wide range of sizes, from small and compact ECs with sizes of $\sim1$ pc to centrally concentrated massive clusters with sizes of $\sim10$ pc - from relatively isolated ECs to large aggregates.

Subsequent early stages of isolated EC formation are seen in Figs.~\ref{f1} and \ref{f2}, respectively.  Isolated small ECs appear to be relatively common in the Galaxy \citep[][Papers I and III]{Camargo11}. Some of the new findings (e.g. C 514 and C 530 in Fig.~\ref{f1}) show clear evidence of ongoing star formation.

\subsection{Composite ECs}

Figs.~\ref{f4} and \ref{f5} show dust complexes where new ECs are born together, forming EC aggregates. Embedded cluster aggregates are groups of clusters close to each
other,  formed together after the gravitational collapse and fragmentation of a GMC or a complex of them. Recently, we reported in Paper III an EC aggregate formed by 7 ECs with similar age
in the Perseus arm. We suggested  that an entire GMC or a cloud complex may fragment almost simultaneously generating EC aggregates. Spiral arms may play an important role in the large-scale structure triggering sequential EC formation by shock compression of large molecular cloud complexes \citep{Fuente08, Camargo11, Camargo12}. EC aggregates are also reported in \citet{Camargo11, Camargo13}.

The shell configuration  of  dust  and the stellar content in Fig.~\ref{f4} point to a sequential EC formation via a \textit{collect and collapse} event \citep{Camargo11}. In the classical \textit{collect and collapse} scenario \citep{Elmegreen77, Whitworth94}  an HII region expands, excited by winds from OB stars, collecting the surrounding material that becomes unstable, and the collapse triggers a sequential star formation. A non-symmetric expanding shell generated by multiple excitation sources or interactions with the surrounding environment may favour the agglutination of neutral material forming dense clumps that are cluster progenitors. The distribution of B stars in Fig.~\ref{f4} possibly created winds that accumulated material in multiple clumps of dust in the surroundings. On the contrary, other B stars may have have ejected primordial gas from an early cluster, stopping the process of star formation. In addition to runaway massive stars, part of the Galactic isolated OB field stars may be related to such failed clusters. If, on one hand, OB type stars may disrupt completely an EC during the primordial gas expulsion, on the other hand, they may form a second-generation of ECs via sequential cluster formation.

\begin{figure}
\resizebox{\hsize}{!}{\includegraphics{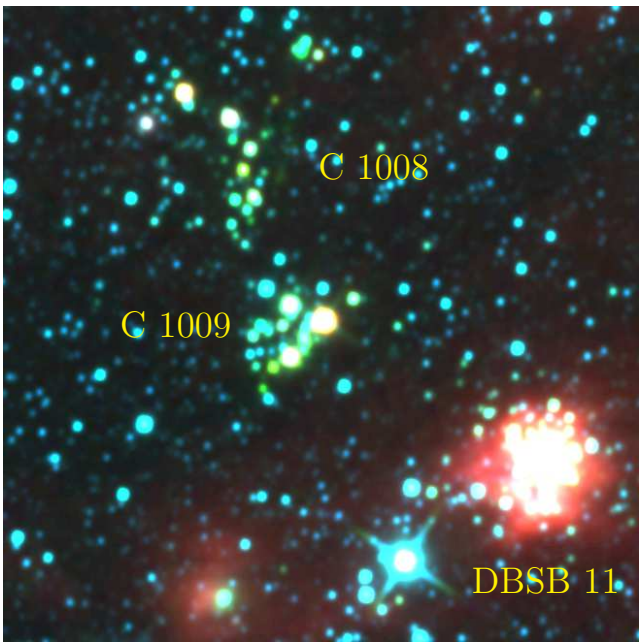}}
\put(-100.0,180.0){\makebox(0.0,0.0)[5]{\fontsize{14}{14}\selectfont \color{yellow}C 1008}}
\put(-175.0,120.0){\makebox(0.0,0.0)[5]{\fontsize{14}{14}\selectfont \color{yellow}C 1009}}
\put(-35.0,23.0){\makebox(0.0,0.0)[5]{\fontsize{14}{14}\selectfont \color{yellow}DBSB 11}}
\caption[]{WISE ($15'\times15'$) RGB image of close multiple clusters C 1008, C 1009 and DBSB 11.}
\label{f6}
\end{figure}

Fig.~\ref{f6} shows multiple closely packed stellar clumps with ongoing star formation. These structures appear to be common in the Galaxy \citep[][Paper I]{Camargo11, Camargo12, Camargo15a}.
Often these types of cluster-forming regions are considered as a single cluster with fractal structure or sub-clusters. On the other hand, in the early evolutionary stages several Galactic isolated ECs present structure with sizes of $\sim1$ pc \citep[][and Paper I]{Testi97, Bica03a, Lada03, Motte03, Adams06, Rodrigues08, Gutermuth09, Higuchi09, Piatti10, Camargo11, Myers11, Alexander12}, similar to those shown in Figs.~\ref{f1} and \ref{f2}. Compact ECs like C 941 (Fig.~\ref{f1}) also appear to be relatively common in the Galaxy. Therefore, the possibility that massive clusters form by merger of several close ECs can not be ruled out.  The fact of considering such systems as  isolated clusters with fractal structure or compact EC aggregates probably will not change their final fate, since both scenarios possibly will lead to a single cluster. However, since in the merging process the ECs do not dissolve, the \textit{infant mortality} may not be as significant as proposed in previous works. As \citet{Lada03} argue that the number of ECs exceeds the number of OCs observed by over an order of magnitude, a high \textit{infant-mortality} rate is assumed for ECs. 

Nevertheless, since  the merging possibility in the early formation is not considered, the EC dissolution rate may be overestimated. \citet{Fellhauer05b} pointed out that within dense aggregates some clusters may merge before the gas expulsion greatly increasing their survival probability. Thus, EC aggregates may be sites of massive cluster formation by merger \citep[][and references therein]{Bastian05, Fellhauer05a}. \citet{Walker15} argue that young massive cluster formation may proceed hierarchically rather than through monolithic collapse, with mass becoming more centrally concentrated as the cluster evolves. In this hierarchical merging scenario, small and dense ECs merge to form larger systems leading to a single centrally concentrated cluster \citep{Camargo11, Camargo12, Fujii12}. In addition, the massive cluster formation by merge explain the excessive age spread for embedded clusters. 
To conclude, we call attention to the possible environmental conditions  where such cluster dynamical evolutionary scenarios may occur more or less frequently,
from closely packed ECs (Fig.~\ref{f6}), poorly populated coeval aggregates (Fig.~\ref{f4}) to rich aggregates of  ECs  in large  dust complexes (Fig.~\ref{f5}).

\section{Concluding remarks}
\label{sec:5}

We communicate the discovery of 652 star clusters and stellar groups using WISE. They were found by one of us (D.C.) and are designated Camargo 447 (C 447) to C 1068. The present list is a followup
of the  446 previous  discoveries in Papers I, II and III. In general  the newly found clusters are  ECs or EGrs.
These objects have probably been neglected in the past, because of the difficulty in detecting  them, especially with 2MASS.
WISE has finally revealed them, in particular in  \citet{Majaess13}, Paper I, and the present study.
Heavily dust-obscured embedded objects may be more easily detected by visual inspections on the WISE Atlas than by  automatic searches of stellar overdensities.
The notion that most stars form in massive clusters may also have contributed to that bias. The initial search resulted in  864 objects, but 212 were discarded because they were  already reported in the literature, or for reasons like probable planetary nebulae, or  spurious objects.
Thus,  $75\%$ of our initial sample was considered to be new star clusters, stellar groups  or candidates.

\begin{figure*}
\resizebox{\hsize}{!}{\includegraphics{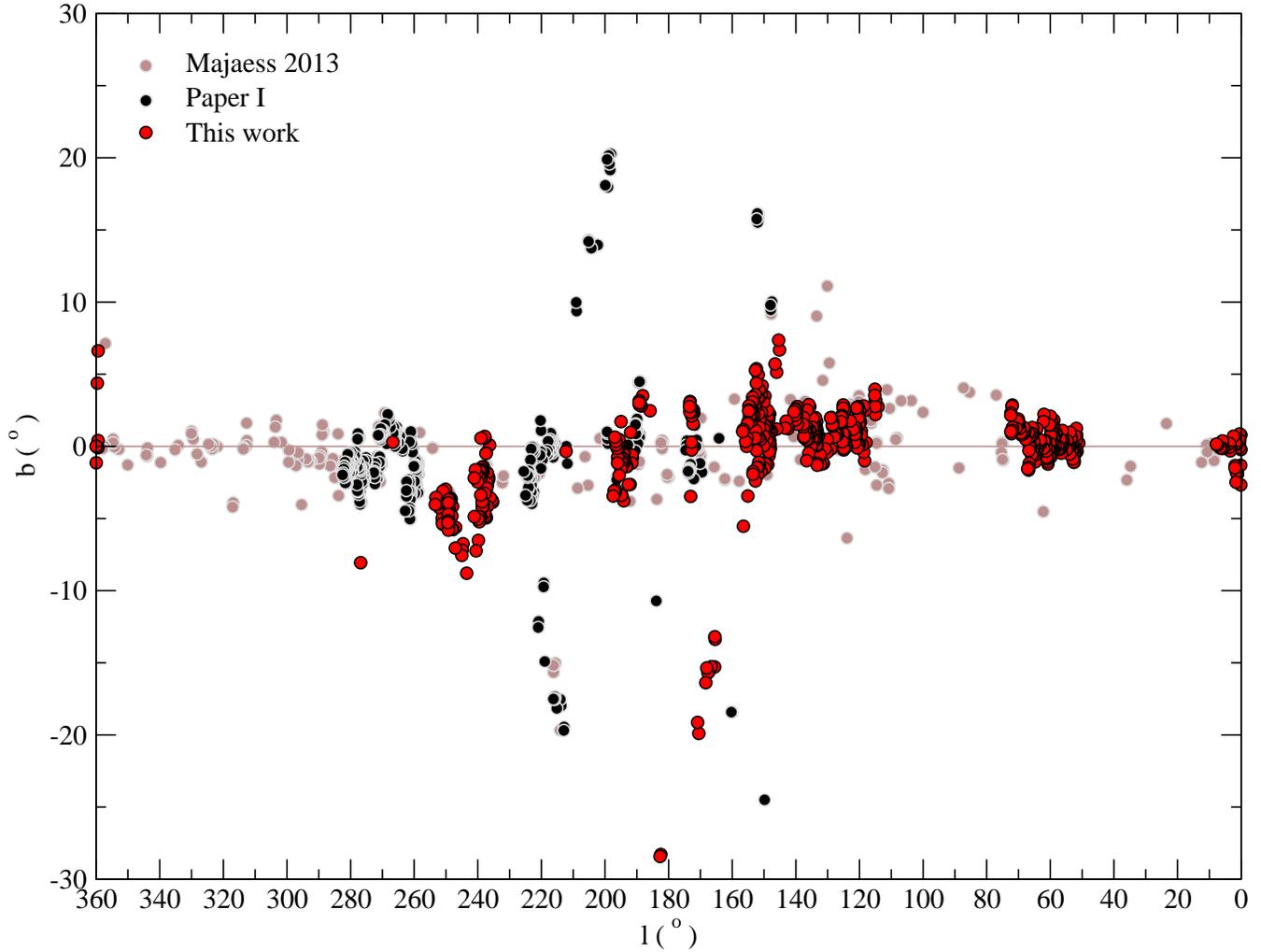}}
\caption[]{Large cluster surveys with WISE: Galactic distribution of the present star clusters and candidates (red points). Comparison with the samples from  Paper I - black points,  and \citet{Majaess13} - brown points. Objects with $|b|<30^{\circ}$ are shown.}
\label{f7}
\end{figure*}

\begin{table*}
\begin{center}
\caption{Derived fundamental parameters for confirmed star clusters in the present study.}
\renewcommand{\tabcolsep}{1.7mm}
\renewcommand{\arraystretch}{1.3}
\begin{tabular}{lrrrrrrrrrr}
\hline
\hline
Cluster&phase&$\alpha(2000)$&$\delta(2000)$&$E(J-H)$&Age&$d_{\odot}$&$R_{GC}$&$x_{GC}$&$y_{GC}$&$z_{GC}$\\
&&(h\,m\,s)&$(^{\circ}\,^{\prime}\,^{\prime\prime})$&(mag)&(Myr)&(kpc)&(kpc)&(kpc)&(kpc)&(pc)\\
($1$)&($2$)&($3$)&($4$)&($5$)&($6$)&($7$)&($8$)&($9$)&($10$)&($11$) \\
\hline
C 716 &EC&2:01:31&60:32:42&$0.50\pm0.03$&$2\pm1$&$3.1\pm0.4$&$9.5\pm0.3$&$-9.26\pm0.29$&$2.31\pm0.33$&$-62.39\pm8.92$\\
C 741 &EC&2:25:39&61:13:33&$0.44\pm0.03$&$2\pm1$&$2.8\pm0.4$&$9.4\pm0.3$&$-9.14\pm0.27$&$1.99\pm0.28$&$19.29\pm2.76$\\
C 758 &EC&2:37:33&61:13:21&$0.32\pm0.01$&$1\pm0.5$&$5.1\pm0.7$&$8.7\pm0.4$&$-7.04\pm0.02$&$0.97\pm0.13$&$-4.99\pm0.69$\\
C 789 &EC&3:15:11&59:54:44&$0.27\pm0.02$&$3\pm1$&$3.1\pm0.3$&$9.8\pm0.2$&$-9.61\pm0.23$&$2.00\pm0.19$&$103.84\pm9.90$\\
C 793 &EC&3:17:23&60:02:07&$0.26\pm0.03$&$4\pm1.5$&$3.0\pm0.4$&$9.7\pm0.3$&$-9.53\pm0.33$&$1.93\pm0.28$&$113.55\pm16.24$\\
C 853 &EC&4:07:32&50:30:48&$0.39\pm0.03$&$2\pm1$&$3.2\pm0.5$&$10.2\pm0.4$&$-10.06\pm0.41$&$1.53\pm0.22$&$-61.90\pm8.85$\\
C 943 &EC&6:15:25&19:01:40&$0.22\pm0.03$&$3\pm2$&$2.9\pm0.4$&$10.1\pm0.4$&$-10.05\pm0.40$&$-0.60\pm0.09$&$48.90\pm6.99$\\
C 978 &EC&6:10:50&12:32:45&$0.20\pm0.03$&$3\pm1$&$3.7\pm0.5$&$10.8\pm0.5$&$-10.78\pm0.5$&$-1.10\pm0.16$&$-201.62\pm28.83$\\
C 1043 &EC&7:26:04&-31:39:24&$0.15\pm0.02$&$4\pm2$&$6.0\pm0.8$&$11.2\pm0.5$&$-9.75\pm0.4$&$-5.40\pm0.76$&$-752.84\pm105.58$\\
\hline
\end{tabular}
\begin{list}{Table Notes.}
\item Col. 2: evolutionary phase - EC means embedded cluster; Col. 3 and 4: Central coordinates; Col. 5: $E(J-H)$ in the cluster central region. Col. 6: age, from 2MASS photometry. Col. 7: distance from the Sun. Col. 8: $R_{GC}$ calculated using $R_{\odot}=7.2$ kpc for the distance of the Sun to the Galactic centre \citep{Bica06}. Cols. 9 - 11: Galactocentric components.
\end{list}
\label{tab2}
\end{center}
\end{table*}

\begin{figure*}
\begin{minipage}[b]{0.328\linewidth}
\includegraphics[width=\textwidth]{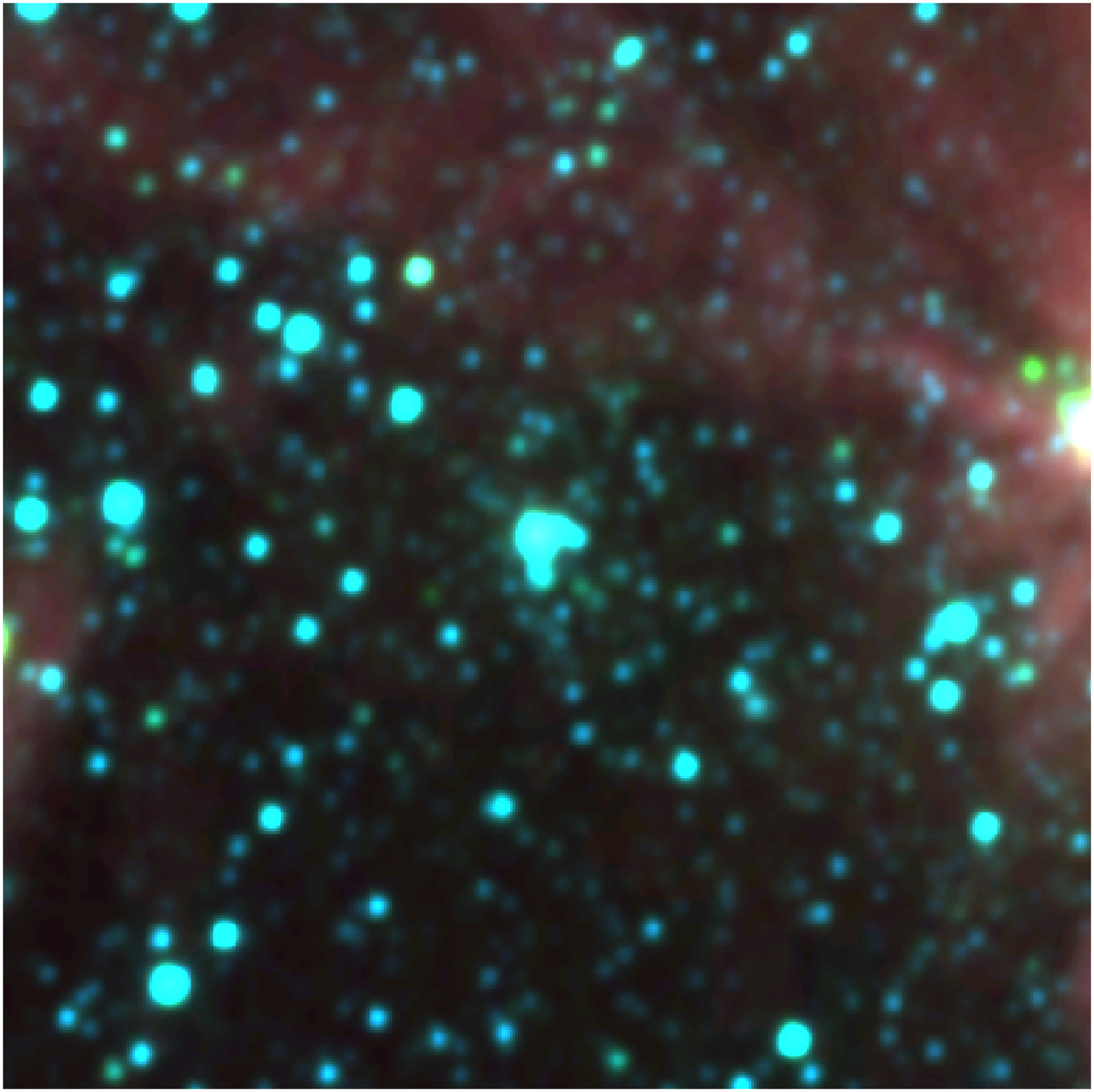}
\put(-120.0,155.0){\makebox(0.0,0.0)[5]{\fontsize{14}{14}\selectfont \color{red} C 943}}
\end{minipage}\hfill
\vspace{0.02cm}
\begin{minipage}[b]{0.328\linewidth}
\includegraphics[width=\textwidth]{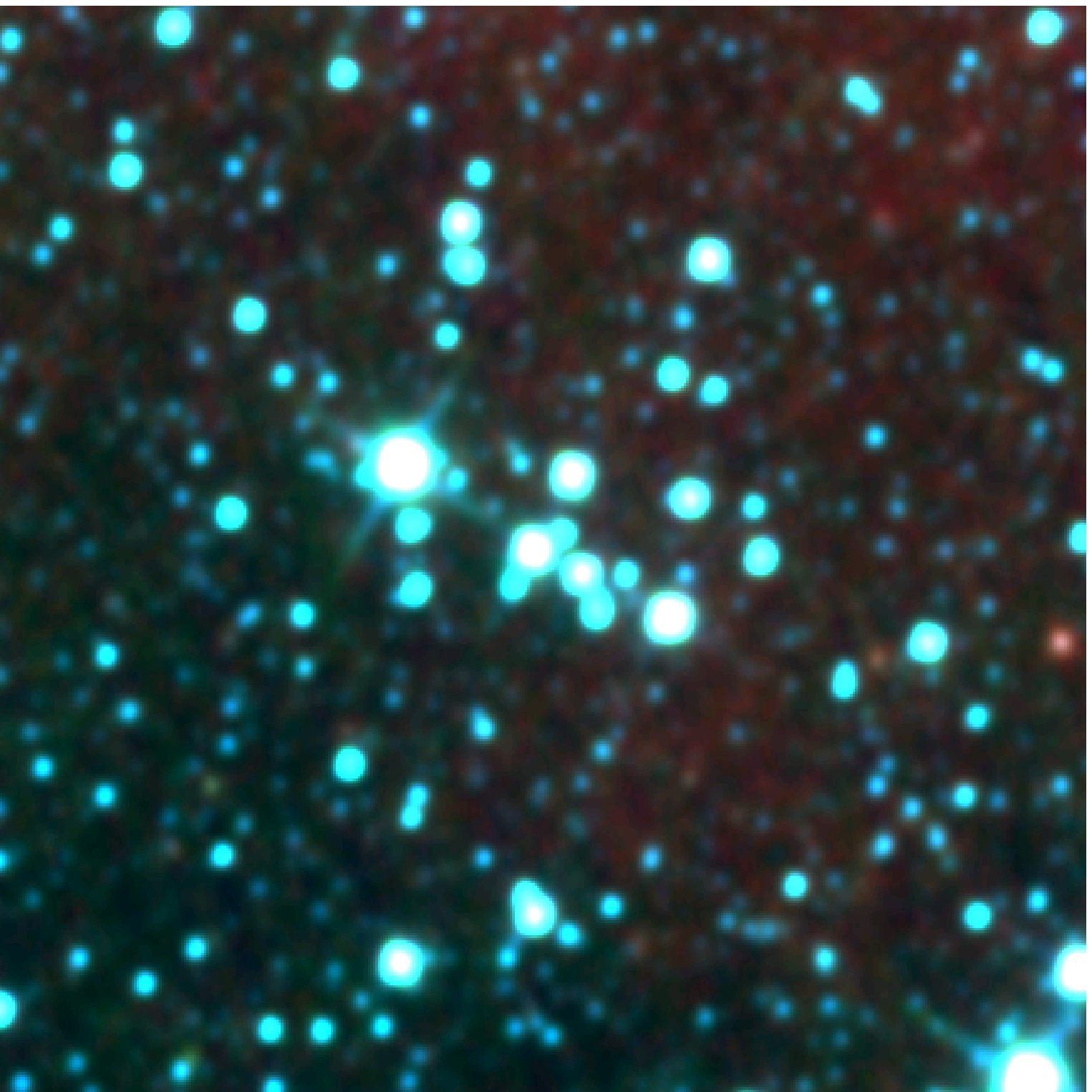}
\put(-120.0,155.0){\makebox(0.0,0.0)[5]{\fontsize{14}{14}\selectfont \color{red}C 1043}}
\end{minipage}\hfill
\vspace{0.02cm}
\begin{minipage}[b]{0.328\linewidth}
\includegraphics[width=\textwidth]{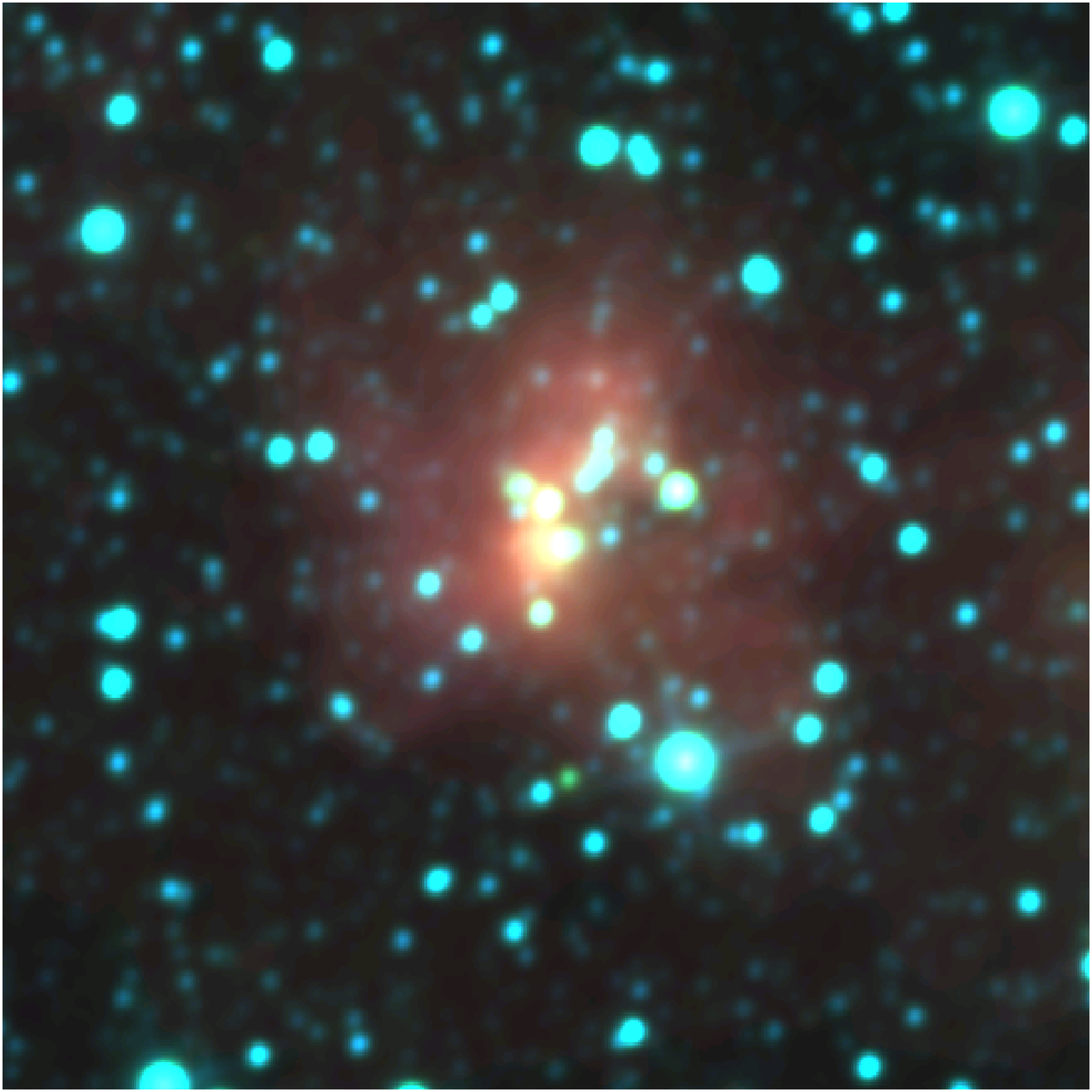}
\put(-120.0,155.0){\makebox(0.0,0.0)[5]{\fontsize{14}{14}\selectfont \color{red} C 716}}
\end{minipage}\hfill
\vspace{0.02cm}
\begin{minipage}[b]{0.328\linewidth}
\includegraphics[width=\textwidth]{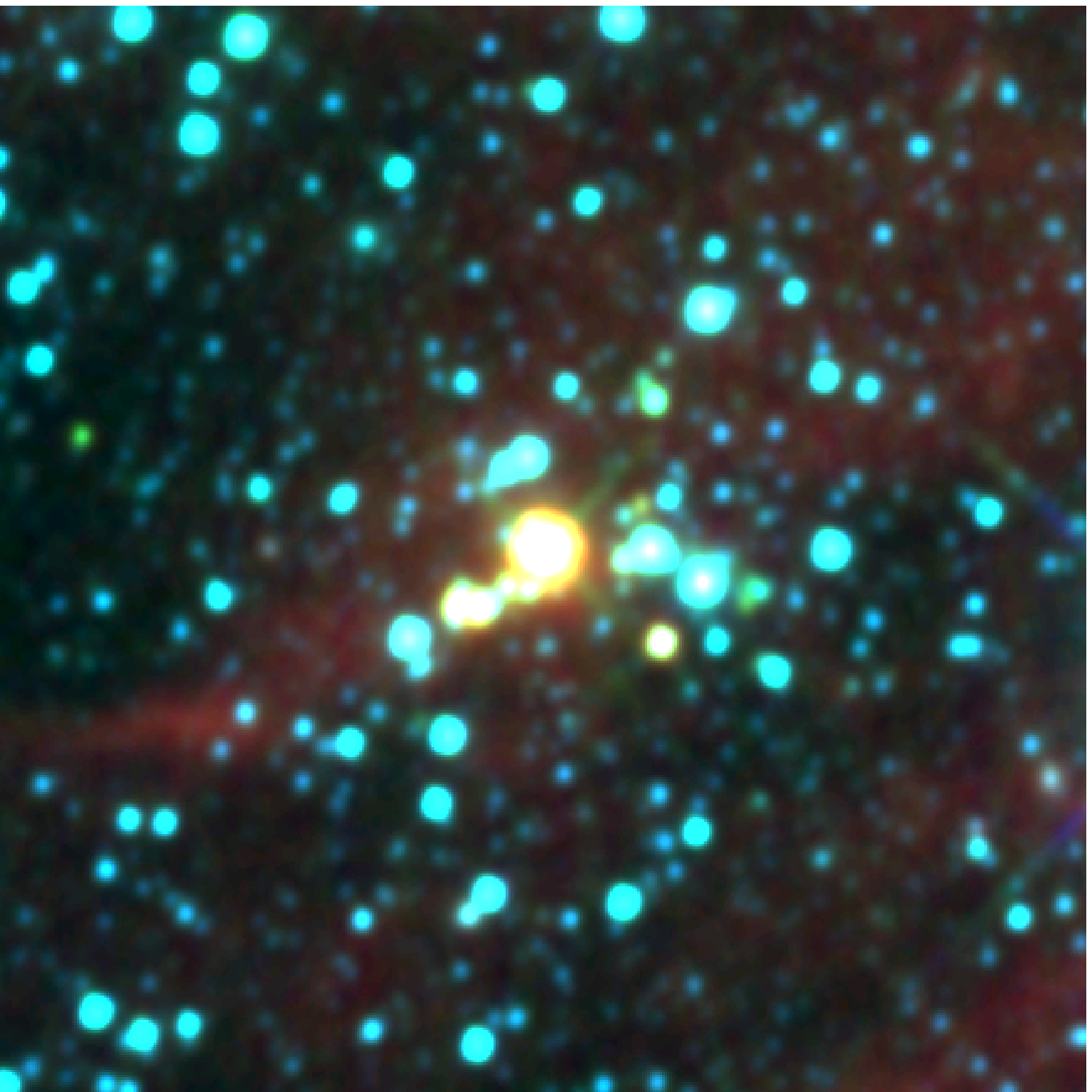}
\put(-120.0,155.0){\makebox(0.0,0.0)[5]{\fontsize{14}{14}\selectfont \color{red}C 978}}
\end{minipage}\hfill
\begin{minipage}[b]{0.328\linewidth}
\includegraphics[width=\textwidth]{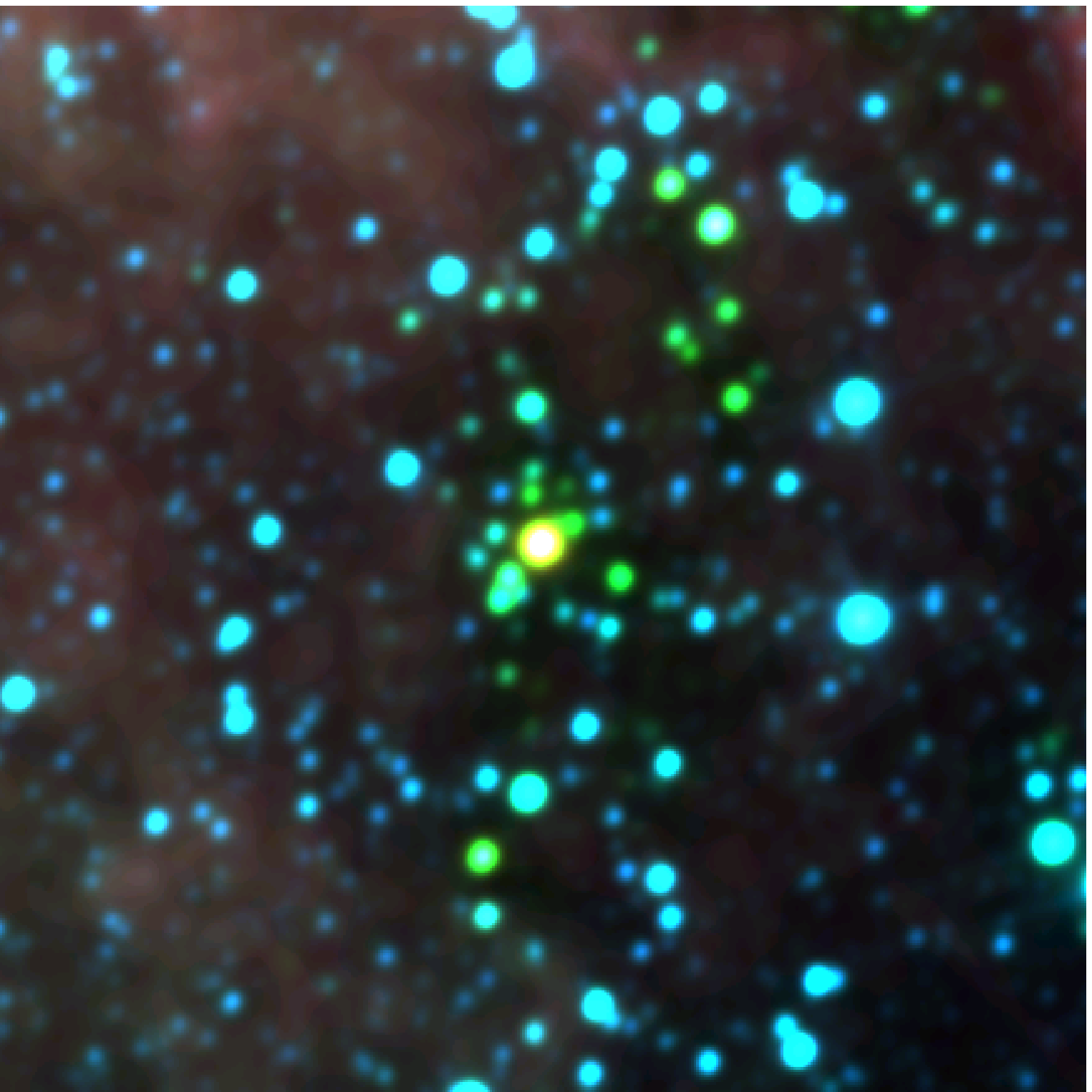}
\put(-120.0,155.0){\makebox(0.0,0.0)[5]{\fontsize{14}{14}\selectfont \color{red}C 741}}
\end{minipage}\hfill
\begin{minipage}[b]{0.328\linewidth}
\includegraphics[width=\textwidth]{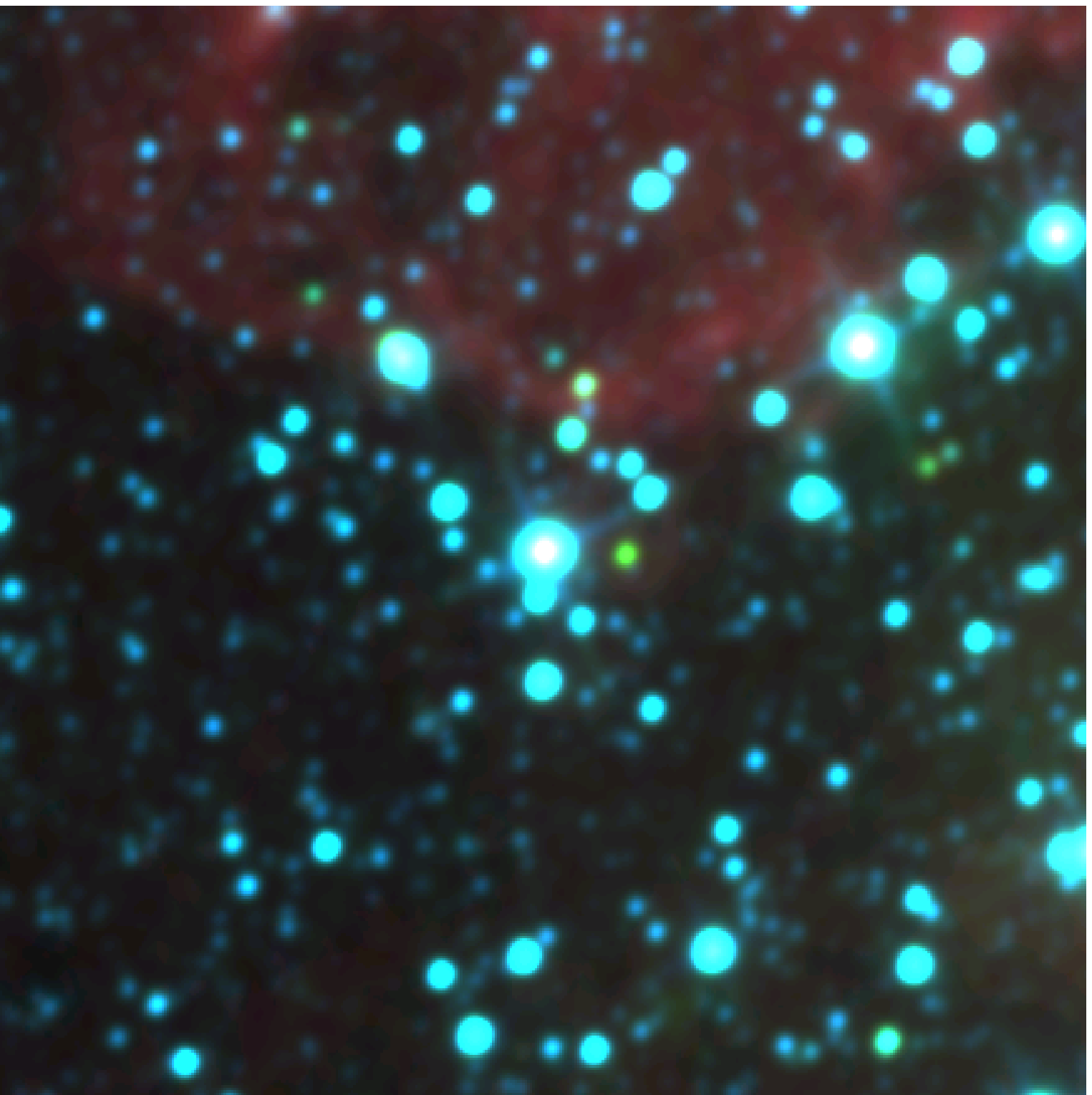}
\put(-120.0,155.0){\makebox(0.0,0.0)[5]{\fontsize{14}{14}\selectfont \color{red}C 793}}
\end{minipage}\hfill
\caption[]{WISE ($10'\times10'$) RGB images centred on the EC C 943, C 1043, C 716, C 978, C 741, and C 793, used in CMD and RDP analyses (Sect.~\ref{sec:3}).}
\label{f8}
\end{figure*}

The discovered embedded clusters provide new insights on  star cluster formation and their early evolution. We point out that several isolated ECs are compact when compared to classical OCs, with sizes comparable to subclusters or fractals in objects often  considered as substructured clusters. On the other hand, some molecular clouds or dust complexes present multiple cluster formation, mainly in filamentary structures. At large scales these structures probably form EC aggregates (e.g. Figs.~\ref{f4} and \ref{f5}).  In small scales they may merge forming a populous (massive) single cluster. The newly found ECs C 1008, C 1009, together with  DBSB 11 (Fig.~\ref{f6}) may have this fate. In this particular viewpoint, some substructured clusters may be only a snapshot view of a merging event. This scenario suggests that there is a continuum of star-forming environments from populous  dense clusters to small ECs in relative isolation, such as C 941. Turbulent GMCs typically develop filamentary structures, which appear to be naturally hierarchical.

\begin{figure}
\resizebox{\hsize}{!}{\includegraphics{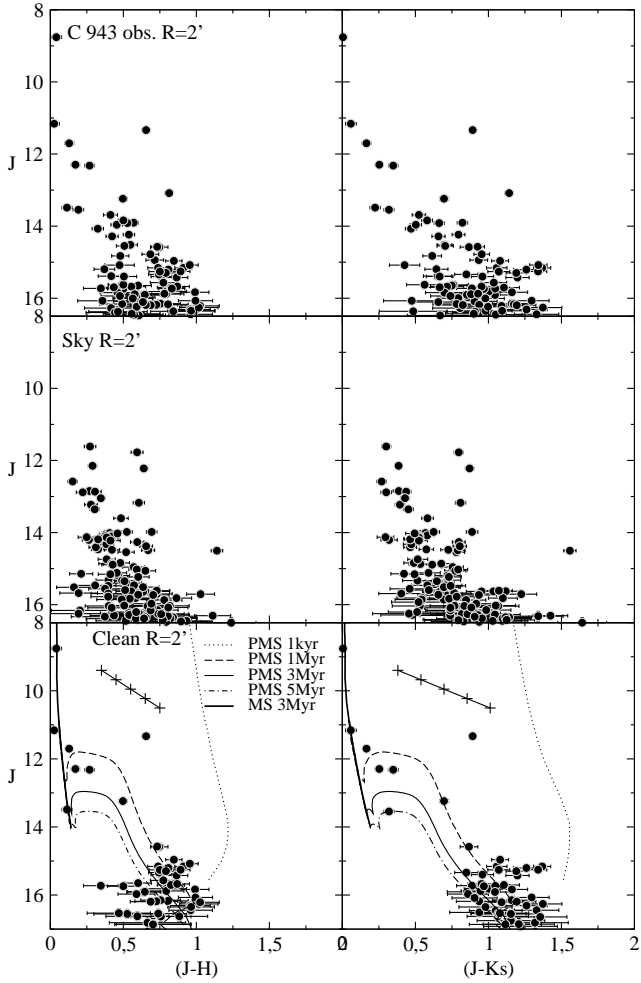}}
\caption[]{2MASS CMDs extracted from the $R=2'$ of C 943. \textit{Top panels}: observed CMDs $J\times(J-H)$ and  $J\times(J-K_s)$. \textit{Middle panels}: equal area comparison field. \textit{Bottom panels}: field star decontaminated CMDs fitted with MS and PMS PARSEC isochrones. We show the reddening vector for $A_V=0$ to 5.}
\label{f9}
\end{figure}

\begin{figure}
\resizebox{\hsize}{!}{\includegraphics{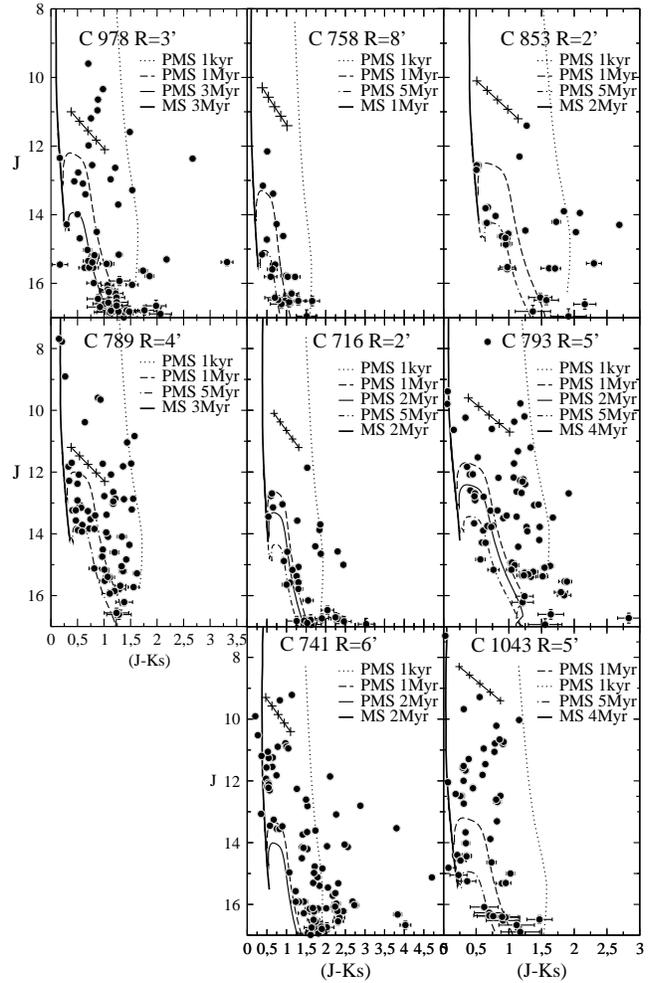}}
\caption[]{2MASS $J\times(J-K_s)$ field star decontaminated Colour-magnitude Diagrams (CMDs) for the clusters in Table~\ref{tab2}. CMDs are fitted with MS and PMS PARSEC isochrones. We show the reddening vector for $A_V=0$ to 5.}
\label{f10}
\end{figure}

We analysed the nature of a representative subsample of 9  clusters (Sect.~\ref{sec:3}), using 2MASS field decontaminated CMDs and RDPs. The inspection of the WISE images (Fig.~\ref{f8}) and 2MASS decontaminated CMDs (Figs.~\ref{f9} and \ref{f10}) confirms them as ECs. Since they are younger than 5 Myr (Table \ref{tab2}), their RDPs behave like those of  many other ECs so far studied with the same method \citep[e.g.][]{Bonatto09}.
They are centrally concentrated, and  have dips owing to dust absorption or crowding. From  their ages they are not expected to follow a King profile, and they do not.

The distribution in Galactic coordinates of the present  sample  (Fig.~\ref{f7}) shows how loci throughout the disc are becoming explored with WISE, as a consequence of  \citet{Majaess13}, Paper I,
and the present study. However,  the plot also points out guiding lines for future searches.

\vspace{0.8cm}

\textit{Acknowledgements}: We thank Dr. B. Twarog for constructive comments and suggestions. This publication makes use of data products from the Two Micron All Sky Survey (2MASS) and Wide-field Infrared Survey Explorer (WISE). The 2MASS is a joint project of the University of Massachusetts and the Infrared Processing and Analysis Centre/California Institute of Technology, funded by the National Aeronautics and Space Administration and the National Science Foundation. WISE is managed and operated by NASA's Jet Propulsion Laboratory (JPL) in Pasadena, California and is a project of the JPL/California Institute of Technology, funded by the National Aeronautics and Space Administration. The spacecraft scanned the entire sky twice. E. Bica and C. Bonatto acknowledge support from CNPq (Brazil).

\onecolumn
\begin{appendix}

  \section{The full list of new star clusters and candidates.}

\footnotesize
\begin{center}
\begin{longtable}[l]{lrrrrrrrr}
\caption{The new list of star clusters or candidates.} \\
\hline \hline
Target&$\ell$&$b$&$\alpha(2000)$&$\delta(2000)$&Size&Type&Comments\\
&$(^{\circ})$&$(^{\circ})$&(h\,m\,s)&$(^{\circ}\,^{\prime}\,^{\prime\prime})$&$(\,^{\prime}\,)$&&\\ \hline
\endfirsthead
\multicolumn{9}{c}{{\tablename} \thetable{} -- Continued} \\
\hline \hline  Target   &  $\ell$   & $b$ & $\alpha(2000)$& $\delta(2000)$   &  Size   & Type  & Comments\\ 
  & $(^{\circ})$   & $(^{\circ})$    & (h\,m\,s) &$(^{\circ}\,^{\prime}\,^{\prime\prime})$    & $(\,^{\prime}\,)$   &     &     \\ \hline 
\endhead
\hline
\multicolumn{9}{l}{{continued on next page\ldots}} \\
\endfoot
\hline \hline
\endlastfoot
Camargo 447 &0.07& -0.22& 17:46:39& -28:59:51&$1.5\times1.5$&EC&\\
Camargo 448 &0.12& -2.67&17:56:26&-30:11:51&$0.7\times0.7$&ECC&\\ 
Camargo 449 &0.09& 0.80&17:42:45&-28:26:24&$3.5\times3.5$&EC&\\
Camargo 450 &0.17& -1.31& 17:51:11& -29:28:14&$    3\times    3$&  ECC&\\
Camargo 451 &0.26& 0.75&  17:43:20& -28:19:26&$  2.5\times  2.5$&  OC&\\
Camargo 452 &0.46& 0.86&  17:43:22& -28:05:47&$  0.8\times  0.8$&  ECC&\\
Camargo 453 &0.47& 0.40&  17:45:12& -28:19:33&$    1\times    1$&  EC&\\ 
Camargo 454 &0.53& 0.18&  17:46:10& -28:23:27&$    6\times    6$&  ECC&      impressive protocluster \\
\end{longtable}
\end{center}
\end{appendix}

\label{lastpage}
\end{document}